\documentclass[10pt.draft]{article}
\usepackage{amssymb}
\usepackage[dvips]{graphicx}
\usepackage{amsmath}
\usepackage{amsfonts}

\input{tcilatex}

\begin{document}

\bigskip

{\LARGE Landau's Last Paper and its Impact on}

{\LARGE Developments in Mathematics, Physics}{\huge \ }

{\LARGE and Other Disciplines in New Millennium}

${\huge \ \ \ \ \ \ \ \ \ \ \ \ \ \ \ \ \ \ \ \ \ \ \ \ \ \ \ \ \ \ \ \ \ }$

\ \ \ \ \ \ \ \ \ \ \ \ \ \ \ \ \ \ \ \ \ \ \ \ \ \ \ A.L. Kholodenko%
\footnote{%
E-mail address: string@clemson.edu}

\textit{375 H.L.Hunter Laboratories, Clemson University, Clemson, }

\textit{SC} 29634-0973, USA

\bigskip

\ 

\bigskip

\textbf{Abstract. }In this paper we discuss the impact of the last Landau
paper on physics and mathematics to date and make some predictions about its
likely impact on sciences in new millennium.

\ 


\ 

PACS (2008): 12.40.-y, 12.60.-i, 11.25.Hf, 12.10.j, 02.10 Kn,

11.25.-w, 05.40.-a

\pagebreak

\ \ \ 

\section{Introduction}

\bigskip

\subsection{General remarks}

\bigskip

The brightest representatives of art, say, Mozart, Mendelsohn, etc., even
though \ they were known as wunderkinds from early age, achieved their full
maturity only at the time of their death, e.g. recall Requiem (for Mozart)
or 9-th symphony for Beethoven, etc. In the case of Landau, his scientific
abilities had become apparent very early and his international acclaim came
to him when he was still in his 30ies. Nevertheless, like with other
greatest, the most profound of his papers were written in a short period of
only 3 years, from 1956 to 1959 (e.g. see the complete list of Landau papers
in this collection) prior to his tragic fatal car accident, which happened
on 7th of January of 1962. As in the case of Mozart or Beethoven, these
works stand in their own classes since, as if anticipating his own end, with
these works he made an attempt to take a look into distant future trying to
foresee contours of the ultimate "theory of everything". In his last paper
"About Fundamental Problems",\ written in 1959 on the occasion of \ Wolfgang
Pauli's death, he summarized \ his own efforts as well as those \ whom he
considered as the best from his generation of physicists. In retrospect,
very much like Einstein who wrote his own obituary at the age of 68, this \
paper happened to become Landau's obituary\ as well.

About 50 years after this paper [1] was written, it is impossible not to be
astonished by Landau's vision of the future, especially, in view of his own
cautious words:

"As is well known, in the case of strong interactions, theoretical physics
at this moment is to a large extent powerless. Therefore, any attempts at
making statements regarding future directions of this field should be
considered as highly speculative (and, hence, prophetic) causing their
authors to be in ever increasing danger of shooting straight into the sky."

In spite of these cautious remarks, he goes on in this paper and makes
"profecies" which happen to be higly accurate. What makes this paper truly
remarkable is the fact that in it (in retrospect) he actually was speaking
on behalf of the whole generation of physicists of his time. The famous
photograph of Dau sitting next to Geo Gamow\ and\ Hendrik Kramers at the
Niels Bohr Institute seminar carries much more than the flavor of these
exciting times. Looking back, one notices that the people sitting in the
front row on this picture: Bohr, Heisenberg, Pauli, Gamov, Landau and
Kramers, although of different age at the time when the picture was
taken-all died within the \ same time span of about 10-15 years. Landau's
paper is an obituary for W. Pauli \ (died in 1958), but then, N.Bohr died in
1962, \ H.Kramers-in 1952, G.Gamov -in 1966, Landau --in 1968 and only
W.Heisenberg died later in 1976. These circumstances make the content of his
last paper especially significant. This is so, because 50 years later upon
its reading, correctness of Landau prophecies is truly remarkable. In the
rest of this paper we would like to explain why this is indeed so and, based
on this, we also will make a few predictions of what lies ahead.

\subsection{Overview of the last Landau paper}

In his obituary, Landau was concerned about the future of high energy
physics leaving aside predictions for other branches of physics. This is
justifiable since historically, especially in Landau's time, the high energy
physics was the ultimate source of inspiration for the rest of physics.
Landau begins with discussion of utility of methods of renormalization and
renormalization group. He notices that in the case of strong interactions
use of the renormalization group methods leads to paradoxical results: even
if one begins with very large couplings at low energies, at high energies
the coupling tends effectively to zero. This phenomenon of "\textsl{%
asymptotic freedom}" was discovered by Pomeranchuk in 1955. \ At the time of
this discovery, to cure such a nulification the \textsl{nonlocality} of
interactions was cautiously suggested. This nonlocality idea was not
welcomed by Landau since, in his opinion, all consequences of the
mathematical apparatus of quantum field theory made without any \ reliance
on a specific model Hamiltonian require locality and are supported by all
available experimental data, in particular, \ by sucsessful use of
dispersion relations. Furthermore, in support of his claim, Landau used the
theory of \textsl{random fragmentation} developed by Fermi [2] for
calculation of angular distribution of pions produced in (high energy)
collisions of two nucleons. It required \ the application of concepts of
statistical thermodynamics to space-time volumes much smaller than those
needed for development of the nonlocal theory. These ideas were checked
experimentally with large degree of sucsess\footnote{%
Deatails will be discussed below.}. Next, Landau notices that Heisenberg [3]
recently expressed the opinion that the existing theory most likely should
undergo substantial changes caused by his S-matrix theory. These changes
however should not be done at the expence of removing the locality
requirement \ in view of the\ assumed validity of dispersion relations
implying Lorentzian causality. Landau goes on and makes a cautious remark
about usefulnes of Heisenberg's nonlinear field theory \ also discussed in
[3]\footnote{%
It should be nothed though that Landau avoids any explicit mentioning about
this nonlinear\ field theory and only remarks that Heisenberg's assumptions
in his opinion are questionable.}. His reservations are based on results of
his work with R.Peierls (done in 1930) which questions the existence of wave
functions in relativistic quantum field theories. It should be said,
however, that such negative result he attributes only to the theory of 
\textsl{strong} interactions. Indeed, he writes: " Operators $\psi $,
containing nonobservable information, should disappear from the theory and,
since the Hamiltonian can be made only from $\psi ^{\prime }s,$ \ we \
necessarily come to the conclusion that \textsl{the} \textsl{Hamiltonian
method for} \textsl{strong interactions outlived itself and should be
burried, of course, with all due respect". }He goes on by saying that the "%
\textsl{basis of new theory should be made out of diagrammatic technique}, 
\textsl{which uses only diagrams with "free" ends, that is typical for
scattering amplitudes and their analytical continuations}. The physical
foundation of such a formalism is made of principles of unitarity and
locality of interactions revealing themselves in analytical properties of
fundamental objects of this new theory, for instance, in all kinds of
dispersion relations \ ..... As result of such an approach, the old problem
of "elementarity" of elementary particles looses its meaning since it cannot
be formulated without interaction between particles." This point of view is
in accord with that made by Heisenberg \ in [3]. The above quotations taken
from Landau paper comprise to a large extent its content.

\section{Impact of Landau paper on high energy physics}

Since the vaue of science for society\ is based on its ability to explain
the present and to predict the future, we would like to apply these criteria
to Landau's last paper. We begin with

\textsl{The} \textsl{theory of random fragmentation} [2]. Development of
this theory resulted in development of QCD and explanation of the phenomenon
of asymptotic freedom\footnote{%
Much more about this theory will be said in Section 5.}. We restore some
details of this development following the recent review paper by Tannenbaum
[4]. In view of this, for the sake of space, we squeeze the number of key
references to the absolute minimum. In the early 60ies (the time when Landau
was incapacitated but still alive) construction of proton accelerators with
energies well above the threshold for anti-proton production, a zoo of new
particles was discovered. Gell-Mann and Ne'eman noticed that particles
sharing the same quantum numbers (spin, parity) follow the symmetry of the
Lie group SU(3). This group is built of three generators which they
associated with three fictitious particles (quarks): u- for "up", \ d-for
"down" and \ s-for "strange"- all having spin 1/2 and fractional charge.
According to the emerging picture mesons are made of two quarks while the
barions-of three.This lead to the discovery of \ a new $\Omega $ particle in
1964 based on predictions of Gell-Mann-Ne'eman theory. This particle had an
apparent problem with the Pauli principle since it was made of three
s-quarks. To avoid this problem, it was suggested in the same year that
quarks come in 3 colours. The major breakthrough came after it was realized
that the SU(3) symmetry is NOT associated with three quarks but with three
colours and that colour-charged gluons are the quanta of asymptotically free
strong interactions which bind hadrons.

\textsl{QCD\ and the asymptotic freedom.} Although the path to asymptotic
freedom begins with Pomeranchuk's paper written in 1955\footnote{%
E.g. see Ref.[2] in Landau's paper [1].},\textbf{\ \ }it\textbf{\ }had many
twists and turns before it reached its final destination culminating in
works by Gross and Wilczek [5] and Politzer [6]. The above mentioned
breaktruth lead to renewed attention to the Yang-Mills (Y-M) theory [7]
which was untill that time in the shadows because of problems with its
renormalizability. The major advancements in solving the renormalization
riddle for the Y-M fields occured in 1967 (when Dau was still alive!). In
this year Faddeev and Popov [8], following earlier general ideas by Dirac
[9], came up with the correct perturbative scheme for the Y-M fields which
takes into account \ gauge constraints and gauge redundancy. The famous
Faddeev-Popov auxiliary fermionic fields (the Faddeev-Popov ghosts) were
incorporated into gauge-invariant renormalization scheme independently by
Slavnov in 1971 [10] and 't Hooft and Veltman \ in 1972 [11]. In the
mean-time using renormalization group methods Bjorken came up with his
famous parton model in 1969 [12] which was sucsessfully tested
experimentally and was instrumental in proving the asymptotic freedom for
the Y-M fields [5,6]. In about the same period Veneziano, following Regge
ideas about the likely form of asymptotics of scattering processes at very
high energies, and in complete accord with Landau's \ predictions \
(regarding the \textsl{central role of amplitudes}) came up with his famous
amplitude in 1968 [13], the year of Landau's death. This amplitude, as is
well known, gave birth to dual resonance models [14] and string theory[15].
Incidentally, much later on the whole diagrammatic machinery of perturbative
calculations in quantum field theories (of whatever kind, in general, and of
QCD in particular) was redone in the string-theoretic fashion [16] making
practical calculations considerably easier. We shall discuss this subject in
some detail in Section 4. In the meantime, we would like to explain why it
was nesessary to come up with string theory in the first place.

\textsl{The standard model}. The story begins with two names: Goldstone and
Higgs. In 1961 Goldstone published his paper [17] on \textsl{global}
symmetry breaking which is always being accompanied by the emergence of a 
\textsl{massless} particle, the "Golstone boson", while Higgs\ in 1964
published a paper on \textsl{local} symmetry breaking causing emergence of
the \textsl{massive} particle subsequestly called \ the Higgs boson [18]. \
In both cases, it is rather easy to recognize Landau's input since the
global symmetry breaking phenomenon follows directly from his work on
phenomenological theory of phase transitions [19], while the breaking of
local symmetry causing emergence of the massive Higgs boson is contained in
his work (with V.Ginzburg) on the phenomenological theory of
superconductivity [20]. Breaking of local symmetry in this case allows to
explain the Meissner effect in superconductors: intitially gauge-invariant
Lagrangian of the Ginzburg-Landau theory of superconductivity (technically
known as Lagrangian for scalar electrodynamics) formally looses its manifest
gauge-invariance since the massless electromagnetic field acquires mass as a
result of spontaneous local gauge symmetry breaking. \ Emergence of this
mass causes the magnetic field to be expelled from the bulk of the
superconductor. \ This is the microscopic cause of the macroscopically
observed Meissner effect. Since many contributions to this centenary volume
discuss this and (related to it) topic(s) in \ sufficient detail, we
continue with other issues in our review.

\ In their Nobel Prize winning lectures both Weinberg and Salam emphasized
the importance of Goldstone and Higgs (and, consequently, Landau's) ideas in
formulation of the unified theory of weak and electromagnetic interactions.
In the Lagrangian \ of the Weinberg-Salam (W-S) model one can easily
recognize the Ginzburg-Landau (superconducting) part responsible for the
emerging Higgs boson. To make any serious calculations with this model
requires use of the renormalization group methods applicable to the
non-Abelian Y-M theories in which gauge invariance is spontaneously broken.
These methods were developed by 't Hooft and Veltman [11]. With help of
these methods, it has become possible to extend the W-S model \ in order to
include strong interactions. The model which unifies weak, electromagnetic
and strong interactions has become known as the \textsl{standard} \textsl{%
model}. Many predictions of this model were thoroughly tested
experimentally. An excellent summary of both theoretical and experimental
results related to the standard model can be found in a recent monograph by
Bardin and Passarino [21]. More relaxed \ and up to date exposition of the
satandard model can be found in [22].

\textsl{Higgs boson and unified field theories}. With all the sucsesses of
the standard \ model, it suffers from two major drawbacks: a) it needs Higgs
(Ginzburg-Landau) field to be renormalizable, b) it does not account for
gravity. Inclusion of gravity into existing unification scheme is
complicated by the well known problems of renormalizability of gravity.
Although, as was demonstrated by Utiyama [23], both gravity and the Y-M
theory are obtainable as result of imposition of the requirement of local
gauge invariance with respect to: a) external (Lorentz) group-for gravity
or, b) with respect to internal (say, isotopic) symmetries in the case of
Y-M, on the underlying field-theoretic Lagrangian, from the point of view of
perturbative renormalization group analysis, gravity escapes treatmets \
which are successful for the Y-M fields. As result, the string theory was
proposed. It is a nonlocal theory, an option which Landau had anticipated
but did not like, and is still in the process of development which, at the
time of this writing, \ has lasted already for 40 years. More on this will
be said in Sections 4 and 5.

As we've mentioned already, the existing string-theoretic formalism can be
adapted to the standard model [16] so that if the string theory is
sucsessful, then the standard model is easily recoverable from it. Having
said this, we silently assumed that the sucsess of string theory depends
very much on its ability to incorporate and to treat gravity on equal
footing with the rest of the forces. This leaves the door totally open for
explanation of the existence of the Higgs boson. In the standard model, even
though it is not yet discovered, its existence is not questioned\footnote{%
Very much the same as the existence of gravitational waves, still to be
discovered. It is interesting to notice that the Google search database has
599,000 entries on Higgs boson and 513,000 entries on gravitational waves.
In both cases one can find many claims that these objects will be
discovered, if not today, then, at worst, by tomorrow morning.} because of
the following: a) many predictions of the standard model not \ directly
involving this boson happen to be correct [21], b) this boson is needed for
this model to be renormalizable, c) only one such boson is sufficient for
the whole machinery to work effectively\footnote{%
Theoretically, nothing forbids us from studying the supersymmetric version
of the Y-M model (and, hence, of QCD and the standard model). Such a model
would require \textsl{more than one} Higgs boson for its renormalizability.
Without discovery of the "standard" Higgs boson such a model thus far is
only of mathematical interest (e.g. read the next section).}. \ The
artificiality of the Higgs boson can be completely removed though if the
already published results contained in our latest works [24,25] become
commonly accepted. In these papers we demonstrated that the \ auxiliary
Ginzburg-Landau (Higgs) functional of the standard model can \ be (\textsl{%
without any approximations }!) rewritten as the Hilbert-Einsten functional
for gravity. This means that such an equivalence converts the \ auxiliary
functional for the Higgs field in the standard model into that for gravity
thus making the already existing theory self-contained\footnote{%
Under such equivalence the Higgs boson \ should be renamed as the
Higgs-Yamabe conformon.}. Such an approach differs substantially from that
recently proposed by Lisi [26] since it does not require any additional
efforts for developing the "theory of everything". It only changes\ our
interpretation of the Higgs boson (very much like the work by Higgs \ which
replaced more narrow\ treatment of the Ginzburg-Landau functional \ taylored
for supercondactivity\ by \ mathematically \ broader tratment enabling to
extract the underlying universality of the Higgs-Landau-Ginzburg phenomenon).

\section{Impact of Landau paper on mathematical physics}

\textsl{Exactly integrable low dimensional \ quantum systems}. Although
Landau was against the nonlinear field theory put forward by Heisenberg [3]
and was advocating instead his diagrammatic methods for amplitudes, his
intuition in this case was incorrect. Subsequent developments revealed that,
actually, \textsl{both} (his and Heisenberg's) \textsl{approaches are
equivalent}. \ Indeed, in 1979 \ the ground breaking paper by Zamolodchikov
brothers was published in Annals of Physics [27]. In it, in complete accord
with Heisenberg, the exact S-matrices were constructed for a number of
manifestly nonlinear relativistic models in 1+1 dimensions. Chronologically,
this work can be considered as a culmination of efforts of many people,
begining with McGuire [28\textbf{]}, in 1964, then Yang [29], in 1967, then
Baxter [30] in 1971, and many many others. The history of discovery of, now
famous, Yang-Baxter (Y-B) equations is well documented in the collection of
reprints assembled by Jimbo [31] where one can find \ as well many
applications of these equations to exactly integrable systems. In fact, the
classical (quantum) system is exactly integrable only if the associated
classical (quantum) Y-B equations can be solved [31,32]. The task of solving
these equations was initiated by Belavin and Drinfel'd [33] and culmunated
in Drinfel'ds paper on Hopf algebras and quantum Y-B equation [34]. An
excellet reference on Hopf algebras, quantum groups, Y-B equations,
Drinfel'd associators and Knizhnik-Zamolodchikov (K-Z) equations is the book
by Kassel [35] where one can also find connections of all these topics with
theory of knots and links. The same topics are discussed in a somewhat more
formal way (with emphasis on symplectic aspects of these issues not present
in Kassel's book) in the monograph by Chari and Pressley [36]. A purely
combinatorial, Heisenberg-style exposition of these topics can be found in
our recent work on Heisenberg's honeycombs [37] in which our readers can
find a very elementary introduction to K-Z equations. These equations were
discovered by Knizhnik and Zamolodchikov [38] in connection with their study
of the Wess-Zumino-Novikov-Witten \ (WZNW) model (the nonlinear sigma model
with the topological WZNW term [39]).Such a model emerged as a by-product of
developments in string theory and in Y-M theory\footnote{%
Where it originated from the Chern-Simons field theory [40,41] which itself
is obtainable from the self-dual Y-M theory to be discussed below.} but is
of independent (from these theories) interest and value in view of its wide
uses in condensed matter [39,42] and conformal field theories [43] where,
incidentally, the Ginzburg-Landau description of such models exists
[24,32,43]. \ In his lecture notes Varchenko argued [44] that all
hypergeometric functions of multiple arguments are solutions of the K-Z
equations. Among these hypergeometric functions of multiple arguments are
all Landau amplitudes as demonstrated by Golubeva [45] and will be discussed
in Section 4. \textsl{Hence, by studying these aplitudes as suggested by} 
\textsl{Landau, one inevitably runs into K-Z equations, quantum groups, Y-B
equations, knots and links, etc. }\ 

\textsl{The} \textsl{Yang-Mills theory in various dimensions and
Ginzburg-Landau model of superconductivity. The role of self-duality}. Most
likely, should he be alive, Landau might argue, based on the knowledge of
his time, that the S-matrices obtained by Zamolodchikov brothers are of
academic interest since they\ only involve nontrivial quantum field theories
in one time and one space dimensions (which is reflected in his skepticism
of the Heisenberg's program). Such a conclusion, however, is not correct
because of the following. On one hand, these quantum models do have
significance in real life in view of their \ experimentally supported
connections with spin chains [39,42]. On another hand, attempts at
quantization of higher dimensional nonlinear quantum fields are most likely
to be in vain\ (in accord with Landau!). The combinatorial arguments in
favour of this conclusion are presented in our recent works [37, 46, 47].
They also follow from Golubeva's paper [45], and will be further explained
in the following section. In support of our point of view, \ we \ also would
like to mention the problem of quantization of pure Y-M field. The
nontriviality of such a quantization already begins in 2 dimensions where
serious study of pure Y-M living on Riemannian surfaces was initiated by
Atiyah and Bott [48]. Their work caused an avalanche of subsequent works by
mathematicians. In particular, in two dimensions, the complete solution of
the pure Y-M field theory was found by Witten [49,50] whose work was
significantly influenced by papers of Atiyah and Bott [48] and Migdal [51].
In dimensions higher than two the situation formally remains much less
tractable since the two-dimentional methods resist generalization to higher
dimensions. Nevertheless, in some special cases, e.g. when the Y-M is
(anti)self-dual, the theory can be developed quite substantially. We
believe, based on the results to be described immediately below, that only
these realizations of the Y-M field make physical sense.

We begin with very significant work by Belavin \textit{et al} [52] done in
1975. In it, the instanton solution of the Euclidean version of the Y-M
field theory was found using essentially ideas of self-duality. In the same
paper the Chern-Simons (C-S) functional was used for the first time in
physics literature\footnote{%
The CS theory was used later on by Witten [40] in his Fields medal winning
paper on the Jones polynomial in which many physical applications are
discussed.}. Shortly thereafter, 't Hooft substantially improved
calculations of Belavin \textit{et al} \ (who essentially obtained only the
leading saddle point result for pure Y-M) by making these calculations more
realistic by incorporating fermions interacting via Y-M fields into these
saddle point-type calculations [53].Atiyah, Hitchin, Drinfeld and Manin
(AHDM instanton) developed \ results by Belavin et al much further using
methods of algebraic geometry.A very readable account of these and related
works can be found in the review paper by Belavin [54]. Donaldson used
essentially these AHDM\ results for development of his theory of topological
classification of four manifolds [55]. In this theory the anti self-dual Y-M
fields play prominent role since only such fields can be associated with
complex structures on 4-manifolds. A very readable account of these
developments including nice overview of the theory of 4 manifolds is given
in a review by Soloviyov [56]. More up to date information, including that
on Seiberg-Witten (S-W) theory, which enabled to simplify considerably the
Donaldson theory is given in the paper by Tyrin [57]. The monograph by
Nicolaescu provides an unprecedented amount of readable material relating
the S-W and Donaldson theories [58]. At the same time, the S-W theory is
just an elaboration on the phenomenological Ginzburg-Landau (G-L) theory of
superconductivity as explained in detail in the monograph by Jost [59]%
\footnote{%
See also the books by \ Pismen [60], where emphasis on physical aspects of
G-L theory is made, and by Yang [61], where mathematical aspects of [60] are
discussed in detail.}. In S-W theory the Abelian gauge field of G-L theory
is replaced by the non-Abelian Y-M gauge field, the boson field is replaced
by the spinor field and the G-L covariant derivative for scalar fields is
replaced by that appropriate for the Dirac fields.

In conclusion, we would like to mention that practically all known exactly
integrable systems can be obtained from the solutions of the (anti)
self-dual Y-M \ field equations \ of Donaldson's theory [62]. Similar
results for the S-W theory are discussed in the monograph by Marshakov
[63].This means that 4-dimensional physics, including gravity (in this case,
the self-dual gravity [64]) can be reduced to the study of two dimensional
problems in the spirit of the \ program outlined by Heisenberg [3] and in
accord with results of our latest works [37,46,47]\footnote{%
More on this is discussed in Sections 4 and 5.}.

\ \textsl{Hitchin equations, Langlands duality, etc}. With exception of the
S-W theory which is an excellent tool for classification of 3 and 4
manifolds but, most likely, cannot be used in its original form for detailed
comparison with experiment, we have not discussed the supersymmetric
versions of Y-M theory. Mathematically, to study such versions is
interesting and leads to issues such as mirror symmetry, Langlands duality,
Hitchin's equations, etc. [65]. However, physically, such theories would
require \textsl{more than one} Higgs boson. Since such a boson remains to be
discovered even in the nonsupersymmetric case\footnote{%
In spite of the very detailed experimental set ups and theoretical
calculations backing these set ups.}, there is no reason to go on \ in this
review with description of ramifications caused by the effects of
supersymmetry (may be, only with the exception of the Faddeev-Popov ghosts
which \ do admit \ supersymmetric treatment [66]\footnote{%
Known as the BRS-type symmetry named after Becchi, Rouet and Stora.} in the
standard Y-M theory).

\section{Impact of Landau paper on mathematics}

\subsection{General remarks}

\bigskip In his paper [1], Landau emphasized the most the results of his
last long and deep paper on analytical properties of the vertex parts
(essentially, the scattering amplitudes) in quantum field theory [67]. This
work can also rightfully to be considered as Landau's last paper. In the
review by Golubeva [45] our readers can find a nice summary of the impact of
this paper on mathematics up to 1976. Accordingly, to avoid duplications, we
shall discuss in this section developments which either happened after this
date or which are not discussed in Golubeva's paper, but are of importance.

In particular, we would like to mention some inconsistency in Landau's
reasoning. On one hand, he wrote his paper [67] in order to make some
progress in the theory of strong interactions. On another, the mathematical
methods based on analysis of Feynman's diagrams which he was employing does
not distinguish between "strong" and "weak interactions. Because of this, we
shall discuss analytical properties of Feynman's \ vertex diagrams for 
\textbf{any} quantum field theory. Using Golubeva's results we are forced to
admit that, whatever this diagram might be, there is an equation for a
hypergeometric-type\ function of multiple arguments whose solution is given
by the vertex part in question. Varchenko demonstrated [44] that all
solutions of hypergeometric equations of multiple arguments are solutions of
the corresponding Knizhnik-Zamolodchikov equations. These solutions are all
\ also expressable in the form the\ Aomoto-Gelfand type hypergeometric
integrals [68-70] \ associated with GKZ equations (also of hypergeometric
type) to be discussed below. Knowledge of analytical properties of the vetex
part allows us to connect it via the optical theorem with the experimentally
observable cross section [71]. Hence, either such experimental data can be
used to reconstruct the vertex part via Kramers-Kronig-type dispersion
relations, or the theoretically obtained vertex part (as some solution of
the K-Z or GKZ equation) can be used for making predictions about the
scattering crossection. In both cases only 2 dimensional results are
actually used in accord with \ remarks made in the preceding section. The
question then emerges: is there a way to classify all solutions of K-Z
equations so that all meaningful scattering amplitudes can be obtained,
recognized and classified? In part, such a classification \ might follow
from works by Belavin and Drinfel'd [33] and Reshetikhin and Wiegmann [72].
And, because of this, all quantum mechanical and field-theoretic models can
be treated by the group-theoretic (which are in the essense topological
[35]) methods involving theory of knots and liks. Such point of view is
developed in our recent paper on Heisenberg honeycombs [37] and has found
its implemetnation in the field-theory case in the works by Connes and
Kreimer [73,74], Connes and Marcolli [75], Kreimer and collaborators [76],
etc. We would like to reobtain some of these results having in mind their
applications in the next section.

Before doing so, it is useful to remind our readers about developments in
mathematics, which were taking place \textsl{exactly} at the time when
Landau's paper [67] was written. Because of these developments, his paper
was immediately appreciated by mathematicians and made a big impact on
mathematics. In 1958, just less than a year before Landau's paper was
written, Rene Thom was awarded the highest award in mathematics-the Fields
Medal-for his contributions to the theory of cobordism [77]. This theory had
played an important role in S-W and Donaldson theories, as well as those
involving Jones polynomial [40], etc. Nevertheless, its is the theory of
singularities and catastrophes developed by Thom between 1954-1962 which is
immediately relevant to Landau's work [67]. That this is indeed the case can
be seen from the monograph by Pham [78] published in 1967. It is essentially
the enlarged version of his earlier publication [79], which, in turn, is
part of a still earlier paper by Fotiadi, Froisaart, Lascoux and Pham
published in the top mathematical journal "Topology" [80]-all devoted to
Landau singularities. Mathemaically, all these works are based on results by
Thom, Leray, Milnor and Lefshetz done in the time period around publication
of Landau's paper. A systematic exposition of these mathematical results can
be found, in addition to works already cited, in the monograph by Arnol'd,
Varchenko and Gussein-Zade [81]. In this book, in addition, one can find the
information about \ the so called \ mixed Hodge structures and period
mapping-results of contributions of Deligne and Griffiths. An excellent
introduction to these topics can be found in the monograph by Carlson,
Muller-Stach and Peters [82]. All these works were employed in our analysis
of analytical properties of Veneziano and Veneziano-Like amplitudes [83-85]
eventually culminating in [47] replacing \ more traditional string-theoretic
models [15] by the tachyon-free model existing in the usual space-time
dimensions. We mention this in view of one of the latest papers by Kreimer
and Bloch [76] in which they noticed relevance of the mixed Hodge structures
to more traditional field theories requiring renormalization. Since
Veneziano postulated (guessed) his amplitudes in 1968 [13] many attempts
were made to derive these amplitudes field-theoretically by studying high
energy asymptotics of scatterring processes where, as is commonly believed,
the scattering amplitudes should exibit the Regge-type behaviour [86,87].
Accordingly, we would like to demonstrate now that, in fact, the Veneziano
amplitudes\footnote{%
We shall call \ scattering amplitudes as "Veneziano amplitudes" \ since they
all can be expressed in terms of, may be, linear combinations of Euler's
beta functions of multiple arguments. The true Veneziano amplitude has in
addition the Regge-type parametrization of these arguments.} are the
backbones of scattering amplitudes of every imaginable field theory. This
fact will be used in the next section in which we are going to demonstrate
much wider uses of Veneziano amplitudes than\ just those in the scattering
processes of high energy physics.

\subsection{ Veneziano amplitudes and \ Feynman diagrams}

We \ begin with the key identity attributed to Feynman [88], page 83, 
\begin{equation}
\frac{1}{\prod\limits_{j=1}^{n}A_{j}^{\lambda _{j}}}=\frac{\Gamma (\lambda
_{0})}{\prod\limits_{j=1}^{n}\Gamma (\lambda _{j})}\prod\limits_{j=1}^{n}%
\left[ \int\limits_{0}^{1}d\alpha _{j}\right] \frac{\delta
(1-\sum\limits_{j=1}^{n}\alpha _{j})\prod\limits_{j=1}^{n}\alpha
_{j}^{\lambda _{j}-1}}{\left( \sum\limits_{j=1}^{n}\alpha _{j}A_{j}\right)
^{\lambda _{0}}}  \tag{4.1}
\end{equation}%
Here $\lambda _{0}=\sum\limits_{j=1}^{n}\lambda _{j}$ and $\Gamma (\lambda
_{j})$ is Euler gamma function. The factors A$_{j}$ are given, as in
Landau's paper[67],\ i.e.as $A_{j}=m_{j}^{2}-k_{j}^{2}$ . In the present
case a slightly more general identity is used\footnote{%
In Landau's paper all $\lambda _{i}^{\prime }s$ are equal to one.}.Such an
identity is employed \ for whatever Feynman integral of the type%
\begin{equation}
I=\int \frac{Bd^{d}p_{1}d^{d}p_{2}\cdot \cdot \cdot }{A_{1}^{\lambda
_{1}}A_{2}^{\lambda _{2}}\cdot \cdot \cdot }  \tag{4.2}
\end{equation}%
with $B$, in the case of fermions, being some polynomial in $k_{i}^{\prime }$%
s with $k_{i}$ being some momenta related to a given line on the diagram.
Clearly, $k_{i}$ may be a combination of both the internal momenta, i.e. $%
p_{i}$, and \ the external momenta, e.g. $l_{i}^{^{\prime }},$ so that a
given diagram is some function of the external momenta and dimensionality $d$
of space (or space-time). Use of (4.1) in (4.2) significantly simplifies
calculations. Details are discussed below.

Looking at (4.1) we notice that in the case if $A_{j}^{\lambda _{j}}=1$ $%
\forall j$ we obtain an important identity 
\begin{equation}
\frac{\Gamma (\lambda _{0})}{\prod\limits_{j=1}^{n}\Gamma (\lambda _{j})}%
\prod\limits_{j=1}^{n}\left[ \int\limits_{0}^{1}d\alpha _{j}\right] \delta
(1-\sum\limits_{j=1}^{n}\alpha _{j})\prod\limits_{j=1}^{n}\alpha
_{j}^{\lambda _{j}-1}=1.  \tag{4.3}
\end{equation}%
This identity defines the Poisson-Dirichlet (P-D) probability measure to be
discussed in some detail in Section 5. Furthermore, without the
normalization factor $\frac{\Gamma (\lambda _{0})}{\prod\limits_{j=1}^{n}%
\Gamma (\lambda _{j})},$ we recognize in the obtained expression the
Veneziano amplitude [83-85]. From this observation the following conclusion
can be drawn: \textsl{All} Feynman's diagrams containing loops
mathematically \textsl{are averages of the stochastic processes of the
Poisson-Dirichlet type\footnote{%
In Section 5 we shall demonstrate that the same is true for traditional
qauantum mechanics and many other\ disciplines.}.}Because of this, all such
diagrams can be rewritten as linear combinations of Veneziano amplitudes.We
would like to provide sufficient evidence that this is indeed the case. This
task is greatly simplified by the significant progress made to date in
actual calculations of Feynman's diagrams. Recent papers by Bogner and
Weinzierl [89] and Weizierl [90] contain a nice summary of these efforts. To
these papers, we would like to add the paper by Tarasov [91] on
hypergeometric representation of a special class of Feynman's diagrams as
well as papers by Smirnov and Smirnov [92] on the use of Gr\"{o}bner bases
for calculation of Feynman integrals. Reading these papers and [16] provides
a broad panorama of efforts made in practical calculations of Feynman
diagrams \ to date, very much in accord with\ and in the spirit of Landau's
predictions.

For the sake of space, and without looss of generality we would like to
consider only the bosonic-type diagrams (that is, the diagrams for which the
factor $B$ in the numerator of (4.2) is identically equal to one). This is
totally justified [89] since the diagrams which contain nontrivial $B$%
-factors can be reduced to those in which $B=1$.In view of this, we are left
with the calculation of integrals of the type

\begin{equation}
I(\mathbf{\alpha ;}s\mathbf{,}t\mathbf{,}u,...)=\int \frac{%
d^{d}p_{1}d^{d}p_{2}\cdot \cdot \cdot }{\left( \sum\limits_{j=1}^{n}\alpha
_{j}A_{j}\right) ^{\lambda _{0}}},  \tag{4.4}
\end{equation}%
where $\emph{\ }\alpha \emph{=}\{\alpha _{1},...,\alpha _{n}\}$ and $%
s,t,u,.. $are some kinematic invariants made of external momenta.
Calculation of such integrals can be found in [89] and, therefore, are of no
immediate concern to us. We shall consider their analytical properties
irresppective to a particular outcome of computation of $I(\mathbf{\alpha ;}s%
\mathbf{,}t\mathbf{,}u,...).$To do so, we need to employ a few facts from
the theory of \ functions of several complex variables. This is needed for
justification of the Laurent-type expansion of $I(\mathbf{\alpha ;}s\mathbf{,%
}t\mathbf{,}u,...).$If such a justification is found, then we can represent $%
I(\mathbf{\alpha ;}s\mathbf{,}t\mathbf{,}u,...)$ as 
\begin{equation}
I(\mathbf{\alpha ;}s\mathbf{,}t\mathbf{,}u,...)=\sum\limits_{l=-\infty
}^{\infty }c_{l}(s,t,u...)\mathbf{\alpha }^{l}  \tag{4.5}
\end{equation}%
with $\ l\mathbf{=[}l_{1},l_{2},...,l_{n}]$; $\mathbf{\alpha }^{l}=\alpha
_{1}^{l_{1}}\cdot \cdot \cdot $ $\alpha _{n}^{l_{n}},$ $l_{1}+l_{2}+..+l_{n}=%
\hat{l};$ $c_{l}(s,t,u...)=c_{l_{1},l_{2},...,l_{n}}(s,t,u,...).$ Clearly,
in actual computations we do not anticipate that we shall use an infinity of
coeffcients c$_{l}$. But, whatever they are, if such expansions \ do exist,
then \textsl{any} Feynman integral is a Poisson-Dirichlet average $<I(%
\mathbf{\alpha ;}s\mathbf{,}t\mathbf{,}u,...)>$ \ with the probability
density defined in (4.3). Hence, the task now lies in providing a
justification that, indeed, such an expansion does exist.

To do so, we need to recall a few facts from the theory of functions of many
complex variables. In particular, we need to recall some models of complex
projective space \textbf{CP}$^{n}.$These are discussed, for instance, in
[93] and were employed in our works [83-85] on Veneziano amplitudes.

Let \textbf{C}$^{n+1}$ be a complex $n+1$ dimensional space and let $\omega
=(\omega _{0},...,\omega _{n})$ be a point in such a space. If we identify
two arbitrary points $\omega ^{\prime }$ and $\omega ^{\prime \prime }$ in 
\textbf{C}$^{n+1}$ via the relation $\omega ^{\prime }=\lambda \omega
^{\prime \prime },$where $\lambda $ is some nonzero complex number, then the
whole space \textbf{C}$^{n+1}$ can be subdivided into equivalence classes
with respect to previously mentioned relation. A quotient \textbf{C}$%
^{n+1}/(\omega ^{\prime }=\lambda \omega ^{\prime \prime })$ is the complex
projective space \textbf{CP}$^{n}.$ Each complex line in \textbf{C}$^{n+1}$
can be characterized by the unit vector $\omega ^{0}$ $=\dfrac{\omega }{%
\left\vert \omega \right\vert }$ . Every \textsl{line} made this way in 
\textbf{C}$^{n+1}$ is a \textsl{point} in \textbf{CP}$^{n}.$Because of this,
we notice that points in \textbf{CP}$^{n}$ are also points in \textbf{C}$%
^{n+1}.$ \ This fact can be used for construction of another model
representing \textbf{CP}$^{n}$. \ For this purpose, let the line in \textbf{C%
}$^{n+1}$ be given parametrically as $z_{\nu }=\omega ^{0}\zeta $, $\nu
=0,...,n$ \ with the parameter $\zeta $ to be determined as follows. If an
equation for a complex sphere $S^{n}$ in \textbf{C}$^{n+1}$ is given by $%
\sum\limits_{i=0}^{n}z_{\nu }\bar{z}_{\nu }=1$, then, we obtain: $\left\vert
\zeta \right\vert ^{2}\sum\limits_{i=0}^{n}\left\vert \omega _{\nu
}^{0}\right\vert ^{2}=1,$ implying $\left\vert \zeta \right\vert ^{2}=1$
which is just an equation for a circle $S.$ \ This means that \textbf{CP}$%
^{n}$ can be determined as quotient of $S^{n}/S.$ Because of this, we can
define the \textsl{deformation retract} of \textbf{CP}$^{n}$ as $%
\sum\limits_{i=0}^{n}t_{i}=1$ by identifying $t_{i}$ with $\left\vert \omega
_{i}^{0}\right\vert ^{2}.$Thus obtained equation is an equation for a $n-$%
simplex $\Delta .$ \ We went into these details only because the $\delta $
constaint in the P-D measure (4.3) \ tells us that the integration in (4.3)
is done over the simplex $\Delta $ and, hence, such an integral lies, in
fact, in the complex projective space (in the case of (4.3), in \textbf{CP}$%
^{n-1})$ in accord with Golubeva's paper [45]. Because of this, the meaning
of auxiliary $\alpha $ variables in (4.5) changes from being real to
becoming complex. This observation provides the needed justification for our
use of the theory of functions of many complex variables. After recognition
of this fact, we have to write down the correct differential form for 
\textbf{CP}$^{n}.$ To do so, we use results of Griffiths discussed both in
Golubeva's paper [45] and in our paper [83] on Veneziano amplitudes, e.g.
see Section 3.1. of [83].Thus, we obtain, 
\begin{equation}
\omega (w)=\sum\limits_{i=0}^{n}(-1)^{i}w_{i}dw[i],  \tag{4.6}
\end{equation}%
where $dw[i]$ = $dw_{1}$ $\wedge \cdot \cdot \cdot \wedge dw_{i-1}\wedge
dw_{i+1}\wedge \cdot \cdot \cdot \wedge dw_{n}.$ Therefore, up to a
constant, any nontrivial Feynman integral involving loops is expressable as 
\begin{equation}
I(s,t,u,..)=\int\limits_{\mathbf{\Gamma }}I(\mathbf{\alpha ;}s\mathbf{,}t%
\mathbf{,}u,...)\prod\limits_{j=1}^{n}\alpha _{j}^{\lambda _{j}-1}\omega
(\alpha )  \tag{4.7}
\end{equation}%
with $\Gamma $ being some homology cycle in \textbf{CP}$^{n}$ (very much
like in the more familiar z-plane case one considers an integral of function 
$f(z)$ along some contour \ which does not cross the singularities of $f(z)$%
) to be determined shortly. For this purpose, we notice that if the Laurent
expansion (4.5) can be used in (4.7), then we end up with the sum of terms
each of which is the Veneziano amplitude. From our work [83] we know than
that each Veneziano amplitude represents some period on the Fermat
hypersurface in \textbf{CP}$^{n}$ 
\begin{equation}
\mathcal{F}(N):x_{0}^{N}+\cdot \cdot \cdot +x_{n}^{N}=0,\text{ }N=1,2,...%
\text{ }  \tag{4.8}
\end{equation}%
so that $\Gamma $ represents one of the periods on such a surface (very much
like in the case of the complex torus \textbf{T} there are two periods).
This fact connects the theory of such integrals with the period mapping
theory, Gauss-Manin connections, hypergeometric functions of many variables,
etc, e.g. see our paper [83] and the original works already mentioned,
e.g.[68,69,81,82]. Such a conclusion depends on the existence of the
Laurent-type expansions of the type given by (4.5) (used in (4.7)).
Fortunately, the existence of such a result also follows, albeit implicitly,
from the Lemma and Theorem in Section 1.4 of Golubeva's paper [45].

We would like now to discuss this topic in some detail. In particular,
following Shabat [93], we notice that upon rescaling, the differential form $%
\omega (w)$ behaves as follows%
\begin{equation}
\omega (w)=(-1)^{\nu -1}w_{\nu }^{n+1}d(\frac{w_{0}}{w_{\nu }})\wedge \cdot
\cdot \cdot \wedge d\left( \frac{w_{\nu -1}}{w_{\nu }}\right) \wedge d\left( 
\frac{w_{\nu +1}}{w_{\nu }}\right) \wedge \cdot \cdot \cdot \wedge d\left( 
\frac{w_{n}}{w_{\nu }}\right) ,  \tag{4.9}
\end{equation}%
implying that a combination $\omega (w)/F(\mathbf{w}),$ with $F(\mathbf{w})$
being some homogenous polynomial of degree $n+1,$ is scale-invariant. This
is the result cited both in our work [83] and in the paper by Golubeva [45].
It belongs to Griffiths [94] who proved that in the projective space \textbf{%
CP}$^{n}$ any closed differential form should look like $\Phi =\frac{P}{F}%
\omega $ with degrees of homogenous polynomials $P$ and $F$ related to each
other as $degP+n+1=degF.$ \ Feynman diagrams discussed in Section 1.6 of
Golubeva's paper are all scale-invariant and obey \ Corollary 2.11. of
Griffiths paper [94].Scale invariance implies renormalizability and vice
versa. In her paper Golubeva talks about convergence of \ Feynman integrals
as a function of their dimensionality without paying attention to the
renormalizability issues. Hence, contrary to \ Landau expectations, the
diagrammatic technique he was advocating is allowing us to look only at the
renormalizable theories. This is not a significant drawback though since, as
we mentioned already, the Landau program also brings to light the
hypergeometric functions of multiple arguments (to be futher discussed
below) and, with them, the Knizhnik-Zamolodchikov equations [38, 44] and,
hence, the WZNW model, etc.The scale-independent integrals are periods of
some varieties, e.g. of the Fermat-type living in the complex projective
space. In accord with general methods developed by Griffiths and nicely
summarized in [82], it is of interest to study these periods as functions of
parameters, in the present case-the parameters are the kinematic invariants.
In addition, as in the theory of ordinary integrals calculated by the method
of residues, in the multidimensional case there is analogous theory
developed by Leray [79,80,93] and is also discussed in Golubeva's paper.
But, as we know from the simpler, one-dimensional case, use of the methods
of residue theory is essentially equivalent to our ability to obtain the
Laurent expansion for a function in question. \ Hence, we come back to the
expansion (4.5).

In the case of\ \ the theory of functions of one complex variable the
procedure for obtaining the Laurent expansion is well known. Such a
procedure is not immediately transferable to the case of many complex
variables though. To do this, one need to ask a question: what is the analog
of the Cauchy formula in the case of many complex variables? Surprisingly,
there are \textsl{many} analogs of this formula in the multivariable case.
Let us recall what is involved in the one-dimensional case. For a domain $D$
with boundary $\partial D$ in $z$-plane and the well behaved function $f(z)$
in this domain we have the Cauchy formula%
\begin{equation}
f(z)=\frac{1}{2\pi i}\int\limits_{\partial D}\frac{f(t)dt}{t-z}.  \tag{4.10}
\end{equation}%
It can be used only for $z\in D$. For $z$ outside $D$ the result is zero
since the function under the integral is holomorphic. With help of (4.10)
the $n-$th derivative of $\ f(z)$ is obtained as 
\begin{equation}
f^{(n)}(z)=\frac{n!}{2\pi i}\int\limits_{\partial D}\frac{f(t)dt}{\left(
t-z\right) ^{n+1}}.  \tag{4.11}
\end{equation}%
This result is used for \textsl{both} the Taylor and \ Laurent expansions of 
$f(z).$ \ By complementarity principle we expect that the same holds true in
the multidimensional case. Hence, we have to find \ a multidimensional
analog \ of (4.10) first. The requirements of scale invariance of integrals
of the type given by (4.7) formally leave us \ with not too much choice. \
Our experience with Veneziano amplitudes [83] tells us, however, that the
situation is considerably trickier. Indeed, taking into account the identity%
\begin{equation}
B(x,y)B(x+y,z)B(x+y+z,u)\cdot \cdot \cdot B(x+y+\cdot \cdot \cdot +t,l)=%
\frac{\Gamma (x)\Gamma (y)\cdot \cdot \cdot \Gamma (l)}{\Gamma (x+y+\cdot
\cdot \cdot +l)},  \tag{4.12}
\end{equation}%
e.g. see equation (3.28) in [83], one can notice that \ study of analytical
properties of any Veneziano amplitude can be reduced to that of the
elementary Euler's beta function $B(x,y)$ which is just a period of the
simplest Fermat variety $\pm z_{0}^{N}+z_{1}^{N}+z_{2}^{N}=0$ so that the
period integral in \textbf{CP}$^{2}$ representing this beta function can be
written as 
\begin{equation}
I=\oint \frac{z_{1}^{c_{1}}z_{2}^{c_{2}}z_{0}^{c_{0}}}{\pm
z_{0}^{N}+z_{1}^{N}+z_{2}^{N}}(\frac{dz_{1}}{z_{1}}\wedge \frac{dz_{2}}{z_{2}%
}-\frac{dz_{0}}{z_{0}}\wedge \frac{dz_{2}}{z_{2}}+\frac{dz_{0}}{z_{0}}\wedge 
\frac{dz_{1}}{z_{1}}),  \tag{4.13}
\end{equation}%
e.g. see (3.10b) of [83]. \ Here $c_{1},c_{2}$\ \ and $c_{0}$ are the same
as x and y in $B(x,y)$ while an auxiliary parameter $c_{0}$ is used only in
the projective form of the beta integral. In actual comutations one should
eventually make a transformation to the affine form where one has to put $%
z_{0}=1$ so that the final result is independent of $c_{0}.$ Deatails can be
found in [83]. For this (projective) period to make sence mathematically the
following arguments should be applied. In the integral one should make the
following replacements: $z_{k}\rightarrow z_{k}\xi ^{j}$, where $\xi
^{j}=\exp (\pm i\frac{2\pi j}{N})$ ($1\leq j\leq N-1)$ and $k=0,1,2$.
Substitution of such an ansatz into (4.13) and requiring that the integral $%
I $ be independent of $\xi ^{j}$ leads to the (Veneziano-type) condition%
\begin{equation}
j_{1}c_{1}+j_{2}c_{2}+j_{0}c_{0}=N  \tag{4.14}
\end{equation}%
with $j_{k}^{\prime }s$ being in the range specified above. Such a condition
makes the period $I$ nonsingular. Physically, however, this case is
uninteresting since the nonsigular expressions cannot be used \ for
scattering amplitudes. The poles in such amplitudes (that is, the
resonances) are being used in the optical theorem for calculation of
experimentally observed cross sections, e.g.read [71]. The way out of this \
apparent difficulty is explained in our work [83], see also our related work
[95]. In view of this, some caution \ should be exercised when one wants to
use the Laurent expansion (4.5) in (4.7). This observation shifts attention
from actual computations of coefficients in the Laurent expansion (4.5)
towards studying of general aspects of such calculations. Our task in this
section lies in proving that \textsl{any} \textsl{scattering amplitude is a
linear combination of the Veneziano-like amplitudes. }This proof is
nonconstructive (that is, it does not allow the effective computation of
amplitudes) but it is essential for the discussion presented in the next
section. Again, we would like to remind our readers that what we call
"Veneziano amplitudes" are in fact Euler's beta functions of multiple
arguments. The true Veneziano amplitudes involve Regge-type parametrization
of these arguments. Analysis made in [83] indicates that in the case when
the Laurent expansion (4.5) has countable infinity of terms with negative
powers, this counable infinity can be squeezed into just one (truly
Veneziano) amplitude using the appropriate Regge-type parametrization. Such
a situation is typical for the meson and hadron physics. In the case when
such an expansion has just a few terms with negative powers, the situation
should be typical for quantum mechanics (see next section), quantum
electrodynamics and weak interactions.

After these \ general remarks, finally, we are ready to provide needed \
(nonconstructive) proof. For this purpose we need only to write the \
multidimensoional analogs of the known Cauchy formulas (4.10), (4.11). In
literature there are three types of \ multidimensional Cauchy-like
formulas-all reduceable to (4.10), e.g. see [93]. We prefer to use the
Martinelli-Bochner (M-B) formula\footnote{%
The more general Cauchy-Fantappie formula discovered by Leray in 1956 is
reducible to that of M-B in the special case $w=\bar{z},$ e.g. read page 158
(after equation 10) of [93]. We are interested just in this special case in
this work.}. It is given by 
\begin{equation}
f(\mathbf{z})=\int\limits_{\partial D}f(\mathbf{\zeta })\omega _{MB}(\mathbf{%
\zeta }-\mathbf{z})  \tag{4.15}
\end{equation}%
with the M-B form $\omega _{MB}$ being defined as 
\begin{equation}
\omega _{MB}(\mathbf{\zeta }-\mathbf{z})=\frac{\left( n-1\right) !}{\left(
2\pi i\right) ^{n}}\sum\limits_{i=1}^{n}\frac{(-1)^{i}(\bar{\zeta}_{i}-\bar{z%
}_{i})d\bar{\zeta}[i]\wedge d\zeta }{\left\vert \mathbf{\zeta }-\mathbf{z}%
\right\vert ^{2n}}  \tag{4.16}
\end{equation}%
and, as usual, $\left\vert z\right\vert ^{2}=\sum\limits_{i=1}^{n}\left\vert
z_{i}\right\vert ^{2}.$ For $n=1$, because of this, we obtain 
\begin{equation}
\omega _{MB}(\zeta -z)=\frac{1}{2\pi i}\frac{\bar{\zeta}-\bar{z}}{\left\vert
\zeta -z\right\vert ^{2}}d\zeta =\frac{1}{2\pi i}\frac{1}{\zeta -z}d\zeta , 
\tag{4.17}
\end{equation}%
as required. The analog of (4.11) in the present case was obtained by
Andreotti and Norguet (e.g.see [93], page 229) and is given by 
\begin{equation}
f^{\left( \mathbf{k}\right) }(\mathbf{z})=\int\limits_{\partial D}f(\mathbf{%
\zeta })\omega _{\mathbf{k}}(\mathbf{\zeta }-\mathbf{z})  \tag{4.18}
\end{equation}%
with the form $\omega _{\mathbf{k}\text{ }}$ being given by 
\begin{equation}
\omega _{\mathbf{k}\text{ }}(\mathbf{\zeta }-\mathbf{z})=\frac{\mathbf{k}%
!\left( n-1\right) !}{\left( 2\pi i\right) ^{n}}\sum\limits_{i=1}^{n}\frac{%
(-1)^{i}(\bar{\zeta}_{i}-\bar{z}_{i})^{k_{i}+1}d\bar{\zeta}^{\mathbf{\alpha }%
+\mathbf{I}}[i]\wedge d\zeta }{\left\vert \mathbf{\zeta }-\mathbf{z}%
\right\vert ^{2n}},  \tag{4.19}
\end{equation}%
where $\mathbf{k}=\{k_{1},...,k_{n}\}$ , $\mathbf{\alpha }=\{\alpha
_{1},...,\alpha _{n}\},\mathbf{I}=\{1,...,1\}$ and $\mathbf{k!=}k_{1}!\cdot
\cdot \cdot k_{n}!.$

Thus, the Laurent expansion \ (4.5) does exists so that all scattering
processes in quantum field and string theory are obtainable as averages of
the stochastic Poisson-Dirichlet type processes. These are going to be
discussed in some detail in the next section. Before doing so, we would like
to conclude this section with a brief discussion of hypergeometric equations
of multiple arguments associated with Feynman diagrams in order to bring
Golubeva's paper [45] up to date.

For the warm up, let us consider a calculation of the following generic
integral%
\begin{equation}
I(z)=\frac{1}{2\pi i}\int\limits_{\gamma }\frac{e^{z\zeta }}{P(\zeta )}d\zeta
\tag{4.20}
\end{equation}%
where $P(z)$ is some poplynomial: $P(z)=z^{n}+a_{1}z^{n-1}+\cdot \cdot \cdot
+a_{n}$ and the contour $\gamma $ is chosen in such a way that all zeros of $%
P(z)$ lie inside. Evidently, 
\begin{equation}
P(\frac{d}{dz})I(z)=0.  \tag{4.21}
\end{equation}%
This equation should be supplemented withthe boudary conditions. They come
from consideration of integrals of the type 
\begin{equation}
I^{\left( k\right) }(z)\mid _{z=0}=\frac{1}{2\pi i}\oint\limits_{\gamma }%
\frac{\zeta ^{k}}{P(\zeta )}d\zeta .  \tag{4.22}
\end{equation}%
Since $k\leq n,$ the function $\dfrac{\zeta ^{k}}{P(\zeta )}$ has at the
point $z=\infty $ zero of order $n-k$. For $k\leq n-2$ we then obtain res$%
_{\zeta =\infty }\dfrac{\zeta ^{k}}{P(\zeta )}=0.$ While for $k=n-1$ we
obtain, $\dfrac{\zeta ^{n-1}}{P(\zeta )}\sim \frac{1}{\zeta }$ for $\zeta
\rightarrow \infty .$Hence, res$_{\zeta =\infty }\dfrac{\zeta ^{n-1}}{%
P(\zeta )}=-1$ Therefore, $I^{\left( n-1\right) }(z)\mid _{z=0}=1$ and the
rest of derivatives are being zero. \ This generic example can be broadly
generalized. This task is accomplished in a series of papers by Gelfand,
Kapranov and Zelevinsky (GKZ) whose results are summarized in [96]. In
addition to this reference, we shall follow in part more relaxed exposition
[97] describing results of GKZ\footnote{%
In our exposition we fill in many gaps in presentation which one might
encounter while trying to read [97].}. For the sake of space, we are not
discussing the related to GKZ issues of \textsl{mirror symmetry}. A very
accessible exposition of this topic is given in our work [95] to which we
refer our readers for details. Reading of this reference should be
sufficient for understanding of finer details of mirror symmetry presented
in [97, 98].

Although not mentioned in the review [96], the original motivation for
sudying the GKZ hypergeometric functions came from observation made by GKZ
in the earlier paper [98] that the integrals of the type 
\begin{equation}
F_{\sigma }(\alpha ,\mathbf{\beta };P)=\int\nolimits_{\sigma
}\prod\limits_{i}P_{i}(x_{1},...,x_{k})^{\alpha _{i}}x_{1}^{\beta _{1}}\cdot
\cdot \cdot x_{k}^{\beta _{k}}dx_{1}\cdot \cdot \cdot dx_{k}  \tag{4.23}
\end{equation}%
with $\alpha =(\alpha _{1},...,\alpha _{m})\in \mathbf{C}^{m},\beta =(\beta
_{1},...,\beta _{k})\in \mathbf{C}^{k},$ $P_{i}(\mathbf{x}%
)=\sum\nolimits_{\omega }v_{\omega }\mathbf{x}^{\omega }$, where $\mathbf{x}%
^{\omega }=x_{1}^{\omega _{1}}\cdot \cdot \cdot x_{k}^{\omega _{k}}$ so that 
$\omega =(\omega _{1},...,\omega _{k})\in \mathbf{Z}^{k}$ and $\sigma $ is
some carefully chosen k-cycle (whose construction is described in [96,
98])-all are integrals of the quantum field theory (qft) already discussed
in Landau's papers [1, 67]\footnote{%
In mathematics these integrals have become known as Aomoto-Gelfand-type
integrals as mentioned already.}! \ Although GKZ promised to write a
separate paper devoted to the qft integrals only, to our knowledge such
paper was never written. Evidently, our equation (4.7) falls into the
category of GKZ integrals as required so not much else can be said. \ To
deal with integrals of the type given in (4.23) consider the following
integral very similar to our (4.7). It is given by 
\begin{equation}
I_{\sigma }^{(m)}(\mathbf{u})=\int\nolimits_{\sigma }P_{\mathbf{u}}(\mathbf{x%
})^{m}\frac{dx_{1}}{x_{1}}\cdot \cdot \cdot \frac{dx_{k}}{x_{k}}  \tag{4.24}
\end{equation}%
Here $m\in \mathbf{Z,}P_{\mathbf{u}}(\mathbf{x})=\sum\limits_{\mathbf{a}\in
A}u_{\mathbf{a}}\mathbf{x}^{\mathbf{a}},$ \textbf{a=(}a$_{1},...,$a$%
_{k}),A=\{$\textbf{a}$_{1},...,$\textbf{a}$_{N}\}$ is some finite subset of 
\textbf{Z}$^{k}$ (usually being \ comprized of vertices of some convex
polytope $\mathcal{P}$ thus leading eventually to the mirror symmetry
arguments as explained in [97, 98]). For simplicity $\sigma =\sigma
_{1}\times \cdot \cdot \cdot \sigma _{k\text{ }}$is a product of $k$ circles 
$\sigma _{i},i=1-k$, $\sigma _{i}\in \mathbf{C,}$ each centered at 0 so that 
$P_{\mathbf{u}}(\mathbf{x})\neq 0$ $\forall $ ($x_{1},...,x_{k})\in \sigma
_{1}\times \cdot \cdot \cdot \sigma _{k\text{ }}.$ By differentiation under
the integral sign \ we obtain%
\begin{equation}
\frac{\partial I_{\sigma }^{(m)}(\mathbf{u})}{\partial u_{i}}%
=m\int\nolimits_{\sigma }\mathbf{x}^{\mathbf{a}}P_{\mathbf{u}}(\mathbf{x}%
)^{m-1}\frac{dx_{1}}{x_{1}}\cdot \cdot \cdot \frac{dx_{k}}{x_{k}}. 
\tag{4.25}
\end{equation}%
To develop this result further some knowledge of solid state physics is
helpful. We would like to remind our readers \ of some useful facts to help
the understanding of what follows. In the case of more familiar 3
dimensional lattices one can choose (rather arbitrarily) a cell made of
vectors \textbf{e}$_{1},\mathbf{e}_{2}$ and \textbf{e}$_{3}$ so that any
vector \textbf{A} of such a 3d lattice is decomposable as \textbf{A}=n$_{1}$%
\textbf{e}$_{1}+$n$_{2}$\textbf{e}$_{2}+$n$_{3}$\textbf{e}$_{3}$ with n$%
_{i}\in \mathbf{Z}$. Suppose now that we translate this lattice as a whole
by vector \textbf{b. }This will cause us to write: \textbf{A}=n$_{1}$\textbf{%
e}$_{1}+$n$_{2}$\textbf{e}$_{2}+$n$_{3}$\textbf{e}$_{3}+\mathbf{b}($n$_{1}+$n%
$_{2}+$n$_{3}).$ \ To make such a vector representation well defined we
mayrequire that n$_{1}+$n$_{2}+$n$_{3}=0.$ If, in addition we fix the
origin, this would result \ in an additional constraint (determining the
location of the origin) n$_{1}$\textbf{e}$_{1}+$n$_{2}$\textbf{e}$_{2}+$n$%
_{3}$\textbf{e}$_{3}=0.$ In the present case we have a lattice $\ $made of
basis set $\{$\textbf{a}$_{1},...,$\textbf{a}$_{N}\}\equiv A,$ so that the
above two conditions are translated into 
\begin{equation}
\mathfrak{l}_{1}+\cdot \cdot \cdot +\mathfrak{l}_{N}=0  \tag{4.26a}
\end{equation}%
and 
\begin{equation}
\mathfrak{l}_{1}\mathbf{a}_{1}+\cdot \cdot \cdot +\mathfrak{l}_{n}\mathbf{a}%
_{N}=0.  \tag{4.26b}
\end{equation}%
Since the vector \textbf{a} in (4.25) belongs to the set A, the condition
(4.26b), when applied to the integral (4.25), leads to the following GKZ
equation%
\begin{equation}
\lbrack \prod\limits_{\mathfrak{l}_{i}<0}\left( \frac{\partial }{\partial
u_{i}}\right) ^{-\mathfrak{l}_{i}}-\prod\limits_{\mathfrak{l}_{i}>0}\left( 
\frac{\partial }{\partial u_{i}}\right) ^{\mathfrak{l}_{i}}]I_{\sigma
}^{(m)}(\mathbf{u})=0  \tag{4.27}
\end{equation}%
This equation is nesessary but is not sufficient for description of\ the
equation for hypergeometric function of multiple arguments since it does not
take into account the homogeneity of such a function. To account for
homogeneity we need to consider what will happen to the integral $I_{\sigma
}^{(m)}(\mathbf{u})$ upon rescaling of the external parameters \textbf{u}.
For this purpose some results from our papers [84, 85] summarized in [100]
are helpful. Omitting technicalities, we would like to present here only the
"bootom line". To do so, we would like to consider again the Laurent
polynomials of the type%
\begin{equation}
f(\mathbf{x})=\sum\limits_{\mathbf{a}\in S_{\sigma }}u_{\mathbf{a}}\mathbf{x}%
^{\mathbf{a}}.  \tag{4.28}
\end{equation}%
Here $S_{\sigma }$ is the polyhedral cone associated with the Newton's
polytope $\mathcal{P}$ . In connection with thus defined $f(\mathbf{x}),$
consider now a general definition of a quasi-homogenous function of degree $%
d $ \ with exponents $\gamma _{1},...\gamma _{n}$. \ Such a function is
defined by%
\begin{equation}
f(s^{\gamma _{1}}x_{1},...,s^{\gamma _{n}}x_{n})=s^{d}f(x_{1},...,x_{n}). 
\tag{4.29}
\end{equation}%
If we apply such a requirement to the individual term \ of the Laurent
polynomial in (4.28), we obtain%
\begin{equation}
\sum\limits_{i=1}^{n}\gamma _{i}\text{a}_{ji}=d_{j}\text{ \ \ }j=1-N\text{.}
\tag{4.30}
\end{equation}%
where N is the number of terms in the Laurent expansion (4.28). Equation
(4.30) is an equation for the hyperplane. Different hyperplanes may have
different $d_{j}^{\prime }s.$ The polytope $\mathcal{P}$ is made of a
collection of \ $N$ hyperplanes. Finally, let us differentiate (4.29) over $%
s $ and let $s=1$ in the end. Thus, we obtain%
\begin{equation}
\sum\limits_{i}\gamma _{i}x_{i}\frac{\partial }{\partial x_{i}}f=df. 
\tag{4.31}
\end{equation}%
At this point our readers already can correctly guess that the required
supplemental equation(s) to (4.27) \ are equations describing the polytope.
In view of (4.30), (4.31), \ these are $k$ equations of the type 
\begin{equation}
\lbrack \sum\limits_{i=1}^{N}\mathbf{a}_{i}u_{i}\frac{\partial }{\partial
u_{i}}-\mathbf{d]}I_{\sigma }^{(m)}(\mathbf{u})=0.  \tag{4.32}
\end{equation}%
We would like to illustrate these general facts by more familiar example of
a usual hypergeometric function. For this, we need to choose the A-system of
basis vectors as follows%
\begin{equation}
A=\left\{ \left[ 
\begin{array}{c}
1 \\ 
1 \\ 
1%
\end{array}%
\right] ,\left[ 
\begin{array}{c}
-1 \\ 
0 \\ 
0%
\end{array}%
\right] ,\left[ 
\begin{array}{c}
0 \\ 
1 \\ 
0%
\end{array}%
\right] ,\left[ 
\begin{array}{c}
0 \\ 
0 \\ 
1%
\end{array}%
\right] \right\} ,  \tag{4.33}
\end{equation}%
while for the \textbf{d}-vector we choose \textbf{d}=(1-c,-a,-b). Under such
circumstances, the GKZ equations can be explicitly written as 
\begin{equation}
\left[ \frac{\partial ^{2}}{\partial u_{1}\partial u_{2}}-\frac{\partial ^{2}%
}{\partial u_{3}\partial u_{4}}\right] \Phi =0  \tag{4.34}
\end{equation}%
\begin{equation}
\left[ u_{1}\frac{\partial }{\partial u_{1}}-u_{2}\frac{\partial }{\partial
u_{2}}\right] \Phi =(1-c)\Phi ,  \tag{4.35}
\end{equation}%
\begin{equation}
\left[ u_{1}\frac{\partial }{\partial u_{1}}+u_{3}\frac{\partial }{\partial
u_{3}}\right] \Phi =-a\Phi ,  \tag{4.36}
\end{equation}%
\begin{equation}
\left[ u_{1}\frac{\partial }{\partial u_{1}}+u_{4}\frac{\partial }{\partial
u_{4}}\right] \Phi =-b\Phi .  \tag{4.37}
\end{equation}%
From the second equation we obtain%
\begin{equation}
\frac{\partial ^{2}}{\partial u_{1}\partial u_{2}}\Phi =u_{2}^{-1}(u_{1}%
\frac{\partial ^{2}}{\partial u_{1}^{2}}+c\frac{\partial }{\partial u_{1}}%
)\Phi  \tag{4.38}
\end{equation}%
From the third and fourth equations we obtain as well%
\begin{equation}
\frac{\partial ^{2}}{\partial u_{3}\partial u_{4}}\Phi
=u_{3}^{-1}u_{4}^{-1}(-u_{1}\frac{\partial }{\partial u_{1}}-a)(-u_{1}\frac{%
\partial }{\partial u_{1}}-b)\Phi .  \tag{4.39}
\end{equation}%
Finally, in view of (4.34) we obtain 
\begin{equation}
u_{3}^{-1}u_{4}^{-1}(u_{1}^{2}\frac{\partial ^{2}}{\partial u_{1}^{2}}%
+(1+a+b)u_{1}\frac{\partial }{\partial u_{1}}+ab)\Phi =u_{2}^{-1}(u_{1}\frac{%
\partial ^{2}}{\partial u_{1}^{2}}+c\frac{\partial }{\partial u_{1}})\Phi . 
\tag{4.40}
\end{equation}%
In this equation we have to set $u_{2}=u_{3}=u_{4}=1$ and $u_{1}=z$ in order
to get the familiar equation for the Gauss hypergeometric function%
\begin{equation}
\left[ z(z-1)\frac{d^{2}}{dz^{2}}+\left[ (a+b+1)z-c\right] +ab\right]
f(a,b,c;z)=0.  \tag{4.41}
\end{equation}%
Recall now that the integral representation for such a function is given by
[101]%
\begin{equation}
f(a,b,c;z)=\frac{\Gamma (c)}{\Gamma (b)\Gamma (c-b)}\int\limits_{0}^{1}dt%
\frac{t^{b-1}(1-t)^{c-b-1}}{(1-tz)^{a}}\equiv <(1-tz)^{-a}>,  \tag{4.42}
\end{equation}%
where the Poisson-Diriclet average $<\cdot \cdot \cdot >$ is defined by 
\begin{equation}
<\cdot \cdot \cdot >=\frac{\Gamma (c)}{\Gamma (b)\Gamma (c-b)}%
\int\limits_{0}^{1}dtt^{b-1}(1-t)^{c-b-1}\cdot \cdot \cdot  \tag{4.43}
\end{equation}%
in accord with (4.3). By the principle of complementarity this means that
all Feynman's diagrams, including those for vertex parts determining the
scattering amplitudes are the P-D averages involving finite (e.g for $a<0$
in (4.42))or infinite (e.g. for $a>0$ in (4.42)) combinations of Veneziano
amplitudes in the sence already described. Other implications of the results
we have just obtained are discussed in the next section.

\ \ \ \ \ \ \ \ \ \ \ \ \ \ \ \ \ \ \ \ \ \ \ \ \ \ \ \ \ \ \ \ \ \ \ \ \ \
\ \ \ \ \ \ \ \ \ \ \ \ \ \ \ \ \ \ \ \ \ \ \ \ \ \ \ \ \ \ \ \ \ \ \ \ \ \
\ \ \ \ \ \ \ \ \ \ \ \ \ \ \ \ \ \ \ \ \ \ \ \ \ \ \ \ \ \ \ \ \ \ \ \ \ \ 

\section{Outlook: Impact of Landau's last paper on sciences in new millenium}

\subsection{General comments}

Although Veneziano had guessed (postulated) his amplitude [13], and some
authors have criticized such an approach to scattering processes of high
energy physics [102], the results of the previous section demonstrate that
Veneziano amplitudes or their linear combinations are intrinsic objects of
high energy scattering processes. The question arises: If this is the case,
what else can be said about these amplitudes? We noticed already in (4.3)
that all such amplitudes are \ P-D averages. Hence, the task now lies in
discussing generalities of such types of stochastic processes. Very
fortunately, without any reference to high energy physics this task was to a
large extent accomplished. References [103-105] provide an excellent
introduction into the theory of the P-D processes which play the central
role in the theory of random fragmentation and coagulation processes. It is
sufficient to type "Poisson-Dirichlet" using Google search engine in order
to find about 52,000 entries. \ Such an abundance of entries is caused by
the fact that many disciplines from the theory of spin glasses to computer
science, from linguistic to forensic science, from economics to population
genetics, from chemical kinetics to random matrix theory, etc.-all involve
the P-D distributions. Remarkably, none of these references mention
applications to the high energy physics, quantum mechanics or qft. The role
of the coagulation fragmentation processes in high energy physics was
recognized some time ago by Mekjian, e.g. see [106] \ and references
therein. His works do not contain, however, arguments and results presented
above, in Section 4, and therefore can be considered as complementary to
ours. We shall say more about this below, in this section. \ 

\subsection{Random fragmentation and coagulation processes and the Dirichlet
distribution}

We begin with some known facts from probability theory. For instance, we
recall that the stationary Maxwell distribution for velocities of particles
in a gas is just of Gaussian-type. \ It can be obtained \ as the stationary
solution of Boltzmann's dynamical equation maximizing Boltzmann's-type
entropy\footnote{%
As discussed recently in our work [25] on the Poincar$e^{\prime }$ and
geometrization conjectures.}. The question arises: Is it possible to find
(discrete or continuous) dynamical equations which will provide known
probability laws as stable stationary solutions? This task will involve
finding of \ dynamical equations along with the corresponding Boltzmann-like
entropies which \ will reach their maxima at respective equilibria for these
dynamical equations. We are certainly not in the position in this closing
section of our paper to discuss this problem in full generality. Instead, we
focus our attention only on processes wihich are described by the so called
Dirichlet distributions. These originate from the integral (equation (2.8)
in \ our work [83] on Veneziano amplitudes) attributed to Dirichlet, that is 
\begin{equation}
\mathcal{D}(x_{1},...,x_{n+1})=\idotsint\limits_{\substack{ u_{1}\geq 0,...,%
\text{ }u_{n}\geq 0  \\ u_{1}+\cdot \cdot \cdot +u_{n}\leq 1}}%
u_{1}^{x_{1}-1}\cdot \cdot \cdot u_{n}^{x_{n}-1}(1-u_{1}-\cdot \cdot \cdot
-u_{n})^{x_{n+1}-1}du_{1}\cdot \cdot \cdot du_{n}.  \tag{5.1}
\end{equation}%
A random vector $(\mathbf{X}_{1},...,\mathbf{X}_{n})\in \mathbf{R}^{n}$ \
such that $\mathbf{X}_{i}\geq 0$ $\forall i$ and $\sum\limits_{i=1}^{n}$%
\textbf{\ \ }$\mathbf{X}_{i}$\textbf{\ }$\mathbf{=}1$\textbf{\ }is said to
be Dirichlet distributed with parameters ($x_{1},...,x_{n};x_{n+1})$ [107]
if the probability density function for $(\mathbf{X}_{1},...,\mathbf{X}_{n})$
is given by 
\begin{eqnarray}
P_{\mathbf{X}_{1},...,\mathbf{X}_{n}}(u_{1},...,u_{n}) &=&\frac{\Gamma
(x_{1}+\cdot \cdot \cdot +x_{n+1})}{\Gamma (x_{1})\cdot \cdot \cdot \Gamma
(x_{n+1})}u_{1}^{x_{1}-1}\cdot \cdot \cdot
u_{n}^{x_{n}-1}(1-\sum\limits_{i=1}^{n}u_{i})^{x_{n+1}-1}  \notag \\
&\equiv &\frac{\Gamma (x_{1}+\cdot \cdot \cdot +x_{n+1})}{\Gamma
(x_{1})\cdot \cdot \cdot \Gamma (x_{n+1})}u_{1}^{x_{1}-1}\cdot \cdot \cdot
u_{n}^{x_{n}-1}u_{n+1}^{x_{n+1}-1},\text{ provided that \ }  \notag \\
u_{n+1} &=&1-u_{1}-\cdot \cdot \cdot -u_{n}.  \TCItag{5.2}
\end{eqnarray}%
To get some feeling of such defined distribution, we notice the following
peculiar aspects of this distribution. For any discrete distribution, we
know that the probability $p_{i}$ must be normalized, that is $%
\sum\nolimits_{i}p_{i}=1.$ Thus, the Dirichlet distribution is dealing with
averaging of the probabilities! Or, better, it is dealing with the problem
of effectively selecting the most optimal probability. The most primitive of
these probabilities is the binomial probability given by 
\begin{equation}
p_{m}=\left( 
\begin{array}{c}
n \\ 
m%
\end{array}%
\right) p^{m}(1-p)^{n-m},\text{ \ }m=0,1,2,....,n\text{.}  \tag{5.3}
\end{equation}%
If $X$ is the random variable subject to this law of probability then, the
expectation \ $E(X)$ is calculated as 
\begin{equation}
E(X)=\sum\limits_{m=1}^{n}mp_{m}=np\equiv \mu .  \tag{5.4}
\end{equation}%
Consider such a distribution in the limit: $n\rightarrow \infty .$ In this
limit, if we write $p=\mu /n$ , then the Poisson distribution is obtained as 
\begin{equation}
p_{m}=\frac{\mu ^{m}}{m!}e^{-\mu }.  \tag{5.5}
\end{equation}%
Next, we notice that $m!=\Gamma (m+1).$ Furthermore, we replace $m$ by real
valued variable $\alpha $ and $\mu $ by $x$. This allows us to introduce the
gamma distribution \ with exponent $\alpha $ whose probability density is 
\begin{equation}
p_{X}(x)=\frac{1}{\Gamma (\alpha )}x^{\alpha -1}e^{-x}  \tag{5.6}
\end{equation}%
for some gamma distributed random variable $X$. \ Based on these results, we
would like to demonstrate now how the Dirichlet distribution can be
represented through gamma distributions. Since the gamma distribution
originates from the Poisson distribution, sometimes in literature the
Dirichlet distribution is called the Poisson-Dirichlet (P-D) distribution
[108]. To demonstrate the connection between the Dirichlet and gamma
distributions is relatively easy. Following Kingman [108], consider a set of
positive independent gamma distributed random variables $Y_{1},...,Y_{n+1}$
with exponents $\alpha _{1},...,\alpha _{n+1}.$ Furthermore, consider the
sum \ $Y=Y_{1}+\cdot \cdot \cdot +Y_{n+1}$ and construct \ a vector $\mathbf{%
u}$ with components: $u_{i}=\frac{Y_{i}}{Y}$. Then, since $%
\sum\nolimits_{i=1}^{n+1}u_{i}$ =$1,$ the components of this vector are
Dirichlet distributed and, in fact, independent of $Y$. \ Details are given
in Appendix A.

Such described Dirichlet distribution is an equilibrium measure in various
fields ranging from spin glasses to computer science, from linguistics to
genetics, from forensic science to economics, etc. [103-105]. Furtheremore,
\ most of fragmentation and coagulation processes involve the P-D
distribution as their equilibrium measure. Some applications of general
theory of these processes to nuclear and particle physics were initiated in
already mentioned series of papers by Mekjian, e.g. see [106]. To avoid
duplications, we would like to rederive some particular results of Mekjian's
differently in order to exibit their connections with the previous section.

\subsection{The Ewens sampling formula and Veneziano amplitudes}

This formula was discussed by Mekjian in [109] without any reference to the
P-D distribution. It is discussed in many other places, including Ewens own
monograph [110]. Our exposition follows work by Watterson [111] where he
considers a simple P-D average of \ monomials of the type generated by the
individual terms in the expansion\footnote{%
Very recently Watterson's results were successfully applied to some problems
in economics [112].} \ 
\begin{equation}
\mathbf{u}^{n}=(u_{1}+\cdot \cdot \cdot
+u_{k})^{n}=\sum\limits_{n=(n_{1},...,n_{k})}\frac{n!}{n_{1}!n_{2}!\cdot
\cdot \cdot n_{k}!}u_{1}^{n_{1}}\cdot \cdot \cdot u_{k}^{n_{k}}.  \tag{5.7}
\end{equation}%
This type of expansion was used in \ our work [83] (equations (2.9),(2.11))
for calculation of multiparticle Veneziano amplitudes. Not surprisingly,
Watterson's calculation also results in the multiparticle Veneziano
amplitude. Upon multiplication by some combinatorial factor in a well
defined limit such an amplitude produces the Ewens sampling formula playing
\ a major role in genetics. Although in Appendix B we reproduce the Ewens
sampling formula (equation (E.6)) without use of the P-D distribution,
Kingman\ [113] demonstrated that "A sequence of populations has the Evens
sampling property if and only if it has the P-D limit" That is to say, the
Ewens sampling formula implies the P-D distribution and vice versa. \textsl{%
In the context of high energy physics it is the same as to say that} \textsl{%
the law of conservation of energy-momentum which must hold for any
scattering amplitude leads to the P-D distribution or, equivalently, to the
Veneziano-type formula for multiparticle amplitudes}. Hence, we expect that
our readers will consult Appendix B prior to reading of what follows.
Furthermore, since the vector \textbf{u} is P-D distributed, it is
appropriate to mention at this point that equation (5.7) represents
genetically the Hardy-Weinberg law [110] for mating species\footnote{%
E.g. see Wikipedia where it is known as Hardy-Weinberg principle.}. Hence,
the Ewens sampling formula provides a refinement of this law accounting for
mutations.

Considear a special case of (5.2) for which $x_{1}=x_{2}=\cdot \cdot \cdot
x_{K+1}=\varepsilon $ and let $\varepsilon =\theta /K$ with parameter $%
\theta $ to be defined later. Then, (5.2) is converted to 
\begin{eqnarray}
P_{\mathbf{X}_{1},...,\mathbf{X}_{K}}(u_{1},...,u_{K}) &\equiv &\phi _{k}(%
\mathbf{u})=\frac{\Gamma ((K+1)\varepsilon )}{\left[ \Gamma (\varepsilon )%
\right] ^{K+1}}\prod\limits_{i=1}^{K+1}u_{i}^{\varepsilon -1}\text{ ,
provided that }  \notag \\
1\text{ } &=&\sum\nolimits_{i=1}^{K+1}u_{i}  \TCItag{5.8}
\end{eqnarray}%
In view of (5.7), let us consider an average $P(n_{1},...,n_{K})$ over the
simplex $\Delta $ (defined by $\sum\nolimits_{i=1}^{K+1}u_{i}=1)$ given by%
\begin{equation}
P(n_{1},...,n_{K})=\frac{n!}{n_{1}!n_{2}!\cdot \cdot \cdot n_{K}!}%
\idotsint\limits_{\Delta }u_{1}^{n_{1}}\cdot \cdot \cdot u_{K}^{n_{K}}\phi
_{k}(\mathbf{u})du_{1}\cdot \cdot \cdot du_{K}.  \tag{5.9}
\end{equation}%
A straightforward calculation produces:%
\begin{equation}
P(n_{1},...,n_{K})=\frac{n!}{n_{1}!n_{2}!\cdot \cdot \cdot n_{K}!}\frac{%
\Gamma ((K+1)\varepsilon )}{\left[ \Gamma (\varepsilon )\right] ^{K+1}}%
\prod\limits_{i=1}^{K}\frac{\Gamma (\varepsilon +n_{i})}{\Gamma
((K+1)\varepsilon +n)}.  \tag{5.10}
\end{equation}%
Up to a prefactor, the obtained product coincides with the multiparticle
Veneziano amplitude discussed in our work [83]. To obtain the Ewens sampling
formula (equation (B.6)) from (5.10) \ a few additional steps are required.
These are: a) we have to let $K\rightarrow \infty $ while allowing many of $%
n_{i}^{\prime }s$ in (5.7) to become zero (this explains the meaning of the
word "sampling"), b) we have to order remaining $n_{i}^{\prime }s$ \ in such
a way that $n_{(1)}\geq n_{(2)}\geq \cdot \cdot \cdot \geq
n_{(k)}>0,0,...,0, $ c) we have to cyclically order the remaining $%
n_{i}^{\prime }s$ in a way explained in Appendix B by introducing $%
c_{i}^{\prime }s$ as numbers of remaining $n_{(i)}^{\prime }s$ \ which are
equal to $i$. That is we have to make a choice between representing $%
r=\sum\nolimits_{i=1}^{k}n_{(i)}$ or $r=\sum\nolimits_{i=1}^{r}ic_{i}$ under
condition that $\ k=\sum\nolimits_{i=1}^{r}c_{i},$ d) finally, just like in
the case of Bose (Fermi) statistics, we have to multipy the r.h.s.of (5.10)
by the obviously looking combinatorial factor $M=K!/[(c_{1}!\cdot \cdot
\cdot c_{r}!)((K-k)!]$. Under such conditions, we obtain: $\Gamma
((K+1)\varepsilon )\simeq \Gamma (\theta ),\Gamma ((K+1)\varepsilon
+r)=\Gamma (\theta +r),\frac{\Gamma (\varepsilon +n_{(i)})}{n_{(i)!}}=\frac{1%
}{n_{(i)}}.$ Less trivial is the result: $K!/[(K-k)!\left[ \Gamma
(\varepsilon )\right] ^{k}]\rightarrow \theta ^{k}.$ Evidently, the factor $%
\dfrac{n!}{n_{1}!n_{2}!\cdot \cdot \cdot n_{K}!}$ \ in (5.10) now should be
replaced by $\dfrac{r!}{n_{(1)}\cdot \cdot \cdot n_{(k)}}.$ Finally, a
moment of thought causes us to replace $n_{(i)}^{\prime }s$ by $\ i^{c_{i}}$%
\footnote{%
This is so because the $c_{i}$ numbers count how many of $n_{(i)}^{\prime }s$
\ are equal to $i$.} in order to arrive at the Ewens sampling formula:%
\begin{equation}
P(k;n_{(1)},...,n_{(k)})=\frac{r!}{[\theta ]^{r}}\prod\limits_{i=1}^{r}\frac{%
\theta ^{c_{i}}}{i^{c_{i}}c_{i}!}  \tag{5.11}
\end{equation}%
in agreement with (B.6). This derivation was made without any reference to
genetics and is completely model-independent. To demonstrate connections
with high energy physics in general and with Veneziano amplitudes in
particular, \ we would like to explain the rationale behind this formula
using an absolute minimum of facts from genetics.

Genetic information is stored in \textsl{genes}. These are some segments (%
\textsl{locuses}) of the double stranded DNA molecule. This fact allows us
to think about the DNA molecule as a world line for mesons made of a pair of
quarks. \ Phenomenologically, the DNA is essentially the \textsl{chromosome}%
. Humans and many other species are \textsl{diploids}. This means that they
need for their reproduction (meiosis) \textsl{two} sets of chromosomes-one
from each parent. Hence, we can think of meiosis as\ a process analogous to
the meson-meson scattering. \ We would like to depict this process
graphically to emphasize the analogy. Before doing so we need to make a few
remarks. First, the life cycle for diploids is rather bizarre. Each cell of
a grown organism contains 2 sets of chromosomes. Maiting, however, requires
this rule to be changed. The \textsl{gametes} (sex cells) from each parent
carry only one set of chromosomes (that is, such cells are \textsl{haploid }%
!). The existence of 2 sets of chromosomes makes individual organism unique
because of the following. Consider, for instance, a specific trait, e.g.
"tall" vs "short". Genetically this property in encoded in some gene%
\footnote{%
Or in many genes, but we talk about a given gene for the sake of argument.}.
A particular realization of the gene (causing the organism to be, say, tall)
is called "\textsl{allele}". Typically, there are 2 alleles -one for each of
the chromosomes in the two chromosome set. For instance, T and t (for "tall"
and "short"), or T and T or t and t or, finally, t and T (sometimes order
matters). Then, if father donates 50\% of T cells and 50\% of t cells and
mother does same, the offspring is likely going to have either TT
composition with probability 1/4, or tt (with probability 1/4) or tT (with
probability 1/4) and, finally, tt with probability 1/4. But, one of the
alleles is usually dominant (say, T) so that we will see 3/4 of tall people
in the offspring and 1/4 short. \ What we just described is the essence of
the Hardy-Weinberg law based, of course, on the original works by Mendel.
Details can be found in genetics literature [110].

Let us concentrate our attention on a particular locus so that the genetic
character(trait) of a particular individual is described by specifying its
two genes at that locus. For $N$ individuals in the population there are $2N$
chromosomes containing such a locus. For each allele, one is interested in
knowing the proportion of $2N$ chromosiomes at which the gene is realized as
this allele. This gives a probability distribution over the set of possible
alleles which describes a genetic make-up of the population (as far as we
are \ only looking at some specific locus). The problem now is to model the 
\textsl{dynamical process} by which this distribution changes in time from
generation to generation accounting for mutations and selection (caused by
the environment). Mutation can be caused just by change of one nucleotide
along the DNA strand\footnote{%
The so called "Single Nucleotide Polymorphism" (SNP) which is detectable
either electrophoretically or by DNA melting experiments, etc.}.Normally,
the mutant allele is independent of its parent since, once the mutation
takes place, it is very unlikely that the corrupt message means anything at
all. Hence, the mutant can be either "good" (fit) or "bad" (unfit) for life
and its contribution can be ignored. If $u$ is the probability of mutation
per gene per generation then, the parameter $\theta =4Nu$ in (6.11). With
this information, we are ready to restore the rest of the genetic content of
Watterson's paper [111]. In particular, random P-D variables $\mathbf{X}_{1},%
\mathbf{X}_{2},...,\mathbf{X}_{K}$ denote the allele relative frequences in
a population consisting of $K$ alleles. Evidently, by construction, they are
Dirichlet-distributed. Let $K\rightarrow \infty $ and let $k$ be an
experimental sample of representative frequencies $k\ll K.$ The composition
of such a sample will be random both, because of the nature of the sampling
process, and because the population itself is subject to random
fluctuations. For this reason we averaged the Hardy-Weinberg distribution
(5.7) over the P-D distribution in order to arrive at the final result
(6.11). This result is an equilibrium result. Its experimental verification
can be found in [110]. It is of interest to arrive at it dynamically. This
is accomplished in the next subsection but in a different context. \ Based
on the facts just discussed, it should be clear that both genetics and
physics of meson scattering (for which Veneziano had proposed his amplitude)
have the same combinatorial origin$.$ All random processes involving
decompositions $r=\sum\nolimits_{i=1}^{k}n_{(i)}$ (or $r=\sum%
\nolimits_{i=1}^{r}ic_{i}$) are the P-D processes [103-105]. \ 

To conclude this subsection, we would like to illustrate graphically why
genetics and physics of meson scattering have many things in common. This is
done with help of the following 3 figures.


\begin{figure}[tbp]
\begin{center}
\includegraphics[width=2.65 in]{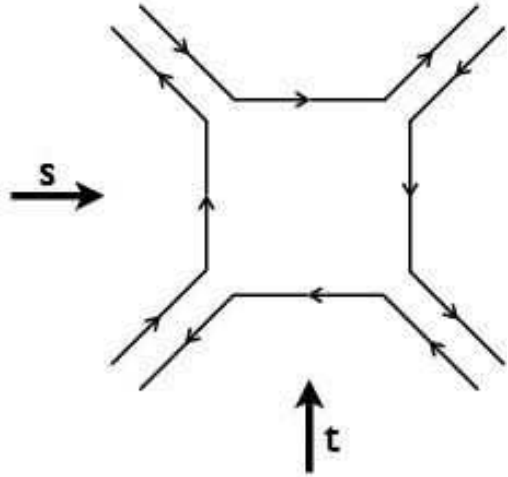}
\end{center}
\caption{The simplest duality diagram describing meson-meson scattering
[114]. The same picture \ describes "collision" of two parental DNA's during
meiosis and can be seen directly under the electron microscope. E.g.see
Fig.2.3 in [115], page 18.}
\end{figure}


\begin{figure}[tbp]
\begin{center}
\includegraphics[width=2.65 in]{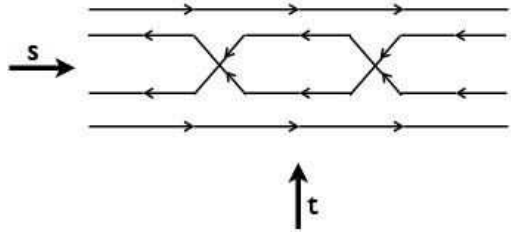}
\end{center}
\caption{Non-planar loop Pomeron diagram for meson-meson scattering [116].
The same diagram describes homologous DNA recombination, e.g. see
Fig.2.2 in [115], page 17.}
\end{figure}


\begin{figure}[tbp]
\begin{center}
\includegraphics[width=2.65 in]{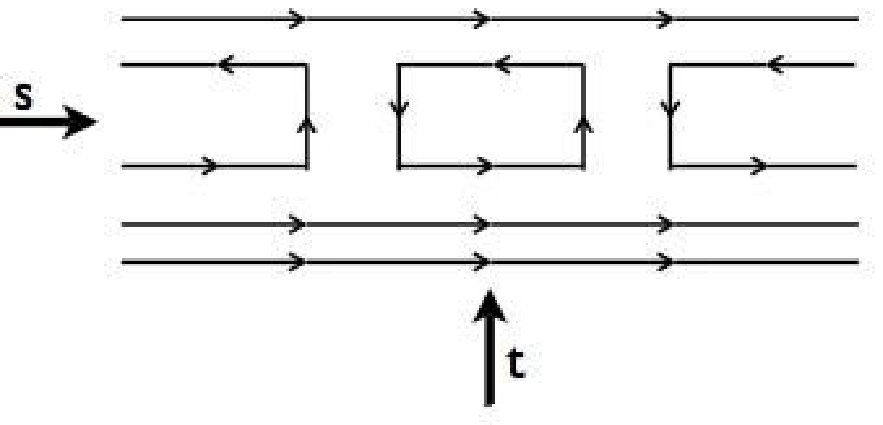}
\end{center}
\caption{The planar loop meson-baryon scattering duality diagram. The same
diagram describes the interaction \ (scattering) between the triple and
double stranded DNA helices [117].}
\end{figure}

\subsection{Stochastic models for \ second order chemical reaction kinetics
involving Veneziano-like amplitudes}

\bigskip

\ The role of stochastic processes in chemical kinetics \ was recognized
long ago. A nice summary is contained in the paper by McQuarrie [118]. The
purpose of this subsection is to connect the results in chemical kinetics
with those in genetics in order to reproduce the Veneziano (or
Veneziano-like) amplitudes as an equilibrium measures for the underlying
chemical/biological processes. Following Darvey \textit{et al} [119] we
consider a chemical reaction $A+B%
\begin{array}{c}
k_{1} \\ 
\rightleftarrows \\ 
k_{-1}%
\end{array}%
C+D$ analogous to the meson-meson scattering processes which triggered the
discovery of \ the Veneziano amplitudes[13]. Let the respective
concentrations of the reagents be $a,b,c$ and $d$. Then, according to rules
of chemical kinetics, we obtain the following "equation of motion"%
\begin{equation}
\frac{da}{dt}=-k_{1}ab+k_{-1}cd.  \tag{5.12}
\end{equation}%
This equation has to be supplemented with the initial condition. It is
obtained by accounting for mass conservation. Specifically, let the initial
concentrations of reagents be respectively: $\alpha =A(0),\beta =B(0),\gamma
=C(0)$ and $\delta =D(0).$ Then, evidently, $\alpha +\beta +\gamma +\delta
=a+b+c+d,$ \ provided that for all times $a\geq 0,b\geq 0,c\geq 0$ and $%
d\geq 0$. Accounting for these facts, equation (5.12) can be rewritten as 
\begin{equation}
\frac{da}{dt}=(k_{-1}-k_{1})a^{2}-[k_{1}(\beta -\alpha )+k_{-1}(2\alpha
+\gamma +\delta )]a+k_{-1}(\alpha +\gamma )(\alpha +\beta ).  \tag{5.13}
\end{equation}%
Thus far, this is a standard result of chemical kinetics. The new element
emerges when one claims that the variables $a,b,c$ and $d$ are random but
are still subject to mass conservation. Then, as we know already from
previous subsections, we are dealing with the P--D-type process. The new
element now lies in the acknowledging the fact that this process is
dynamical. Following Kingman [120] we would like to formulate it in precise
mathematical terms. For this purpose, we introduce the vector \textbf{p}%
(t)=(p$_{1}$(t),..., p$_{k}$(t)) such that it moves randomly on the simplex $%
\Delta $ defined by%
\begin{equation}
\Delta =\{\mathbf{p}(t);p_{j}\geq 0,\sum\nolimits_{i=1}^{k}p_{i}=1\} 
\tag{5.14}
\end{equation}%
In the present case the possible states of the system at time $t$ which
could lead to a new state specified by $a,b,c,d$ at time $t+\Delta t$
involving not more than one transformation in the time interval $\Delta t$
are [119]%
\begin{equation}
\left( 
\begin{array}{cccc}
a+1 & b+1 & c-1 & d-1 \\ 
a-1 & b-1 & c+1 & d+1 \\ 
a & b & c & d%
\end{array}%
\right) .  \tag{5.15}
\end{equation}%
In writing this matrix, following [119], we have assumed that random
variables $a,b,c$ and $d$ are integers.\ Using (5.15) we obtain the
following equation of motion%
\begin{eqnarray}
P(a,b,c,d;t+\Delta t)-P(a,b,c,d;t) &=&[k_{1}(a+1)(b+1)P(a+1,b+1,c-1,d-1;t) 
\notag \\
&&+k_{-1}(c+1)(d+1)P(a-1,b-1,c+1,d+1;t)  \notag \\
&&-(k_{1}ab+k_{-1}cd)P(a,b,c,d;t)]\Delta t+O(\Delta t^{2}).  \TCItag{5.16}
\end{eqnarray}%
In view of the fact that the motion is taking place on the simplex $\Delta ,$
it is sufficient to look at the stochastic dynamics of just one variable,
say, $a$ (very much like in the deterministic equation (5.13)). This
replaces (5.16) by the following result:%
\begin{eqnarray}
\frac{d}{dt}P_{a}(t) &=&k_{1}[(a+1)(a+1+\beta -\alpha
)P_{a+1}(t)+k_{-1}[(\gamma +\alpha -a+1)(\delta +\alpha -a+1)P_{a-1}(t) 
\notag \\
&&-[k_{1}a(\beta -\alpha +a)+k_{-1}(\gamma +\alpha -a)(\delta +\alpha
-a)]P_{a}(t);\text{ \ provided that}  \notag \\
\text{\ \ }P_{\alpha }(0) &=&1,\text{ \ \ }\alpha =a\text{ and }P_{\alpha
}(0)=0\text{ if }a\neq \alpha .  \TCItag{5.17}
\end{eqnarray}%
To solve this equation we introduce the generating function $G(x,t)$ via%
\begin{equation*}
G(x,t)=\sum\limits_{a=0}P_{a}(t)x^{a}
\end{equation*}%
and use this function in (5.17) to obtain the following Fokker--Plank-type
equation%
\begin{eqnarray}
\frac{\partial }{\partial t}G(x,t) &=&x(1-x)(k_{1}-xk_{-1})\frac{\partial
^{2}}{\partial x^{2}}G+(1-x)[k_{1}(\beta -\alpha +1)  \notag \\
&&+k_{-1}(2\alpha +\gamma +\delta -1)x]\frac{\partial }{\partial x}G  \notag
\\
&&-k_{-1}(\alpha +\gamma )(\alpha +\delta )(1-x)G(x,t).  \TCItag{5.18}
\end{eqnarray}%
This equation admits separation of variables: $G(x,t)=S(x)T(t)$ with
solution for $T(t)$ in the expected form: $T(t)=exp(-\lambda _{n}k_{1}t),$
leading to the equation for $S(x)$%
\begin{equation}
x(1-x)(1-Kx)\frac{d^{2}}{dx^{2}}S(x)+[\beta -\alpha +1+K(2\alpha +\gamma
+\delta -1)x](1-x)\frac{d}{dx}S-[K(\alpha +\gamma )(\alpha +\beta
)(1-s)-\lambda _{n}]S(x)=0  \tag{5.19}
\end{equation}%
This equation is of Lame-type as discussed in [44,47] and, therefore, its
solution should be a polynomial in $x$ of degree at most $\varpi ,$ where $%
\varpi $ should be equal to the minimum of ($\alpha +\gamma ,\alpha +\delta
,\beta +\gamma ,\delta +\delta ).$ As in quantum \ mechanics, this implies
that the spectrum of eigenvalues $\lambda _{n}$ is \ discrete, finite and, a
priory nondegenerate. Among these eigenvalues there must be $\lambda _{0}=0$
since such an eigenvalue corresponds to the time-independent solution of
(5.19) typical for the true equilibrium. Hence, for this case we obtain,
instead of (5.19), the following final result:%
\begin{equation}
x(1-Kx)\frac{d^{2}}{dx^{2}}S(x)+[\beta -\alpha +1+K(2\alpha +\gamma +\delta
-1)x]\frac{d}{dx}S-[K(\alpha +\gamma )(\alpha +\beta )]S=0,  \tag{5.20}
\end{equation}%
where $K=k_{-1}/k_{1}.$ This constant can be eliminated from (5.20) if we
rescale $x:x\rightarrow Kx.$ After this, equation (5.20) acquires the
already familiar (e.g. see (4.41)) hypergeometric form%
\begin{equation}
x(1-x)\frac{d^{2}}{dx^{2}}S(x)+[\beta -\alpha +1+(2\alpha +\gamma +\delta
-1)x]\frac{d}{dx}S(x)-(\alpha +\gamma )(\alpha +\beta )S(x)=0.  \tag{5.21}
\end{equation}%
In [120] Kingman obtained the Fokker-Planck type equation analogous to our
(5.18) describing the dynamical process whose stable equilibrium is
described by (5.21) (naturally, with different coefficients) and leads to
the P-D distribution (5.2) essential for obtaining the Ewens sampling
formula. Instead of reproducing his results in this work, we would like to
connect them with results of our Section 4. \ For this purpose, we begin
with the following observation.

\subsubsection{Quantum mechanics, hypergeometric\ functions and P-D
distribution}

In our works [37,46] we provided detailed explanation of the fact that all \
exactly solvable 2-body quantum mechanical problems involve different kinds
of special functions obtainable from the Gauss hypergeometric funcftion
whose integral representation is given by 
\begin{equation}
F(a,b,c;z)=\frac{\Gamma (c)}{\Gamma (b)\Gamma (c-b)}\int%
\limits_{0}^{1}t^{b-1}(1-t)^{c-b-1}(1-zt)^{-a}dt.  \tag{5.22}
\end{equation}%
As is well known from quantum mechanics, in the case of a disctete spectrum
all quantum mechanical problems involve orthogonal polynomials. The question
then arises: under what conditions on coefficients ($a,b$ and $c$) \ can the
infinite hypergeometric series whose integral representation is given by
(5.22) be reduced to a finite polynomial? This happens, for instance, if we
impose the \textsl{quantization condition}: $-a=0,1,2,....$ In such a case
we can write $(1-zt)^{-a}=\sum%
\nolimits_{i=1}^{-a}(_{i}^{-a})(-1)^{i}(zt)^{i} $ \ and use this finite
expansion in (5.22). In view of (5.2) we obtain the convergent generating
function for the Dirichlet distribution (5.2). Hence, \textsl{all \ known
quantum mechanical} \textsl{problems involving discrete spectrum} \textsl{%
are examples of the P-D stochasic processes\footnote{%
Evidently, in the case of continuum spectrum we are also dealing with the
P-D processes but the corresponding hypergeometric series conain now
countable infinity of terms.}}. \textsl{For hypergeometric functions of
multiple arguments this was demonstrated in Section 4}. \ \textsl{Thus, all
quantum mechanical, quantum field-theoretic} \textsl{and string-theoretic
processes are the P-D processes}. As such they fall into\textbf{\ a much
larger class} \textbf{of stochastic processes known as random} \textbf{%
coagulation and fragmentation processes}. We would like to conclude this
section with the following additional observations.

\subsubsection{Hypergeometric functions, Kummer series expansions and
Veneziano amplitudes}

In view of just introduced quantization condition, the question arises: is
this the only condition reducing the hypergeometric function to a polynomial
? More broadly: what conditions on coefficients $a,b$ and $c$ \ should be
imposed so that the function $F(a,b,c;z)$ becomes a polynomial?\footnote{%
Incidentally, in the case of K-Z-type hypergeometric equations such a
problem was solved only in 2007 [121]!} The answer to this question was
provided by Kummer in the first half of 19th century [101]. We would like to
summarize his results and to connect them with results presented above.
Additional details can be found in our paper [47]. By doing so we shall
reobtain Veneziano amplitudes for chemical process described by equation
(5.21).

According to general theory of hypergeometric equations [101], the infinite
series for a hypergeometric function degenerates to a polynomial if one of
the numbers%
\begin{equation}
a,b,c-a\text{ or \ }c-b  \tag{5.23}
\end{equation}%
is an integer. This condition is equivalent to the condition that, at least
one of the eight numbers $\pm (c-1)\pm (a-b)\pm (a+b-c)$ is an odd number.
This \ observation produces 24 solutions for the Gauss hypergeometric
function (5.21) found by Kummer. Among these he singled out 6 (generating
all 24) and among these 6 he established that every 3 of them are related to
each other.

Let us denote these 6 functions (solutions) as $u_{1},...,u_{6}$ . Then, we
can represent, say, $u_{2}$ and $u_{6}$ using $u_{1}$ and $u_{5}$ as the
basis set. We can do the same with $u_{1}$ and $u_{5}$ by representing them
through $u_{2}$ and $u_{6}$ and, finally, we can connect $u_{3}$ and $u_{4}$
with $u_{1}$ and $u_{5}.$ Hence, for our purposes, it is sufficient to
consider, say, $u_{2}$ and $u_{6}.$ We obtain, 
\begin{equation}
\left( 
\begin{array}{c}
u_{2} \\ 
u_{6}%
\end{array}%
\right) =\left( 
\begin{array}{cc}
M_{1}^{1} & M_{2}^{1} \\ 
M_{1}^{2} & M_{2}^{2}%
\end{array}%
\right) \left( 
\begin{array}{c}
u_{1} \\ 
u_{5}%
\end{array}%
\right) ,  \tag{5.24}
\end{equation}%
with $M_{1}^{1}=\dfrac{\Gamma (a+b-c+1)\Gamma (1-c)}{\Gamma (a+1-c)\Gamma
(b-c+1)};$ $M_{2}^{1}=\dfrac{\Gamma (a+b+1-c)\Gamma (c-1)}{\Gamma (a)\Gamma
(b)};M_{1}^{2}=\dfrac{\Gamma (c+1-a-b)\Gamma (1-c)}{\Gamma (1-a)\Gamma (1-b)}%
;M_{2}^{2}=\dfrac{\Gamma (c+1-a-b)\Gamma (c-1)}{\Gamma (c-a)\Gamma (c-b)}.$
The determinant of this matrix becomes zero if either two rows or two
columns become the same\footnote{%
The condition for the determinant to become zero is the resonance condition.
It is of central importance in both quntum mechanics and string theory [47].}%
. For instance, we obtain:%
\begin{equation}
\dfrac{\Gamma (a)\Gamma (b)}{\Gamma (c-1)}=\dfrac{\Gamma (a-c+1)\Gamma
(b-c+1)}{\Gamma (1-c)}\text{ and }\frac{\Gamma (c-a)\Gamma (c-b)}{\Gamma
(c-1)}=\frac{\Gamma (1-a)\Gamma (1-b)}{\Gamma (1-c)}.  \tag{5.25}
\end{equation}%
For $c=1$ we obtain an identity. From [119] we find that (5.21) admits 2
independent solutions:%
\begin{equation}
S(x)=\left\{ {}\right. \frac{\text{either }F(-\alpha -\gamma ,-\alpha
-\delta ,\beta -\alpha +1;Kx),\text{ for }\beta \geq \alpha \text{ }}{\text{%
or }\left( Kx\right) ^{\alpha -\beta }F(-\beta -\gamma ,-\beta -\delta
,\alpha -\beta +1;Kx),\text{ for }\beta \leq \alpha \text{ }}.  \tag{5.26}
\end{equation}%
Hence, the condition $c=1$ \ in (5.25) causes two solutions for $S(x)$ to
degenerate into one polynomial solution, provided that we make an
identification: $\beta =\alpha $ in (5.26). Notice that \ to obtain this
result there is no need to impose an extra condition: $a=b\footnote{%
Here $a$ and $b$ have the same meaning as in (5.22) and should not be
confused with concentrations.}$ (which, in our case, is the same as $\gamma
=\delta ).$

This makes sence physically both in chemistry and in high energy physics. In
the case of high energy physics, if the Veneziano amplitudes are used for\
description of, say, $\pi \pi $ scattering, in [83], page 54,\ it is
demonstrated that processes for which "concentrations "$a=b$ cause this
amplitude to vanish. The Veneziano condition: $a+b+c=-1($ equation $(1.5)$
of [83]) has its analog in chemistry where it plays the same role, e.g. of
mass conservation. In the present case we have $\alpha +\beta +\gamma
+\delta =const,$ and the Veneziano-like amplitude obtainable from
(5.25),(5.26) is given now by 
\begin{equation}
V_{c}(a,b)=\frac{\Gamma (-\alpha -\gamma )\Gamma (-\alpha -\delta )}{%
-c\Gamma (-c)}\mid _{c=1}.  \tag{5.27}
\end{equation}%
In view of known symmetry of the hypergeometric function: $%
F(a,b,c;x)=F(b,a,c;x)$, we also have: $V_{c}(b,a)=V_{c}(a,b).$ This is
compatible with the symmetry for Veneziano amplitude. To make the analogy
with Veneziano amplitudes complete, we have \ to select the following
options in (5.27): a) $\alpha =0,\gamma =1,\delta =1,2,...$or$;$ $b)$ $%
\alpha =1,\gamma =0,\delta =0,1,2,...$ These conditions are exactly
compatible with those in (1.19) of [\textbf{83}] for Veneziano amplitudes.
Finally, in view of (5.22), these are the quantization conditions needed for
resonances to exist as required.

\section{Conclusions}

In Einstein's time to create a unified field theory was the ultimate goal of
physics. It should be noted though, that, for instance, Pauli was not
sharing Einstein's beleif in the unified field theory. Abdus Salam in his
Nobel Lecture given in 1979\footnote{%
Readily available at the Google search database.} on pages 518-519 writes: "
I must admit I was taken aback by Pauli's fierce prejudice against
universalism-against what we would today call unification of basic
forces-but I did not take it too seriously. I felt this was a legacy of the
exasperation which Pauli had always felt at Einstein's somewhat formalistic
attempts at unifying gravity with electromagnetism- forces which in Pauli's
phrase "cannot be joined-for God hath rent them asunder". Based on the
results discussed in this paper, we sincerely hope, that should Pauli \ be
alive, and be aware of the enormous universality of the stochastic theory of
random fragmentation-coagulation processes, he would change his position on
universalism. \ By writing an obituary for Pauli, Landau, in fact, had
inagurated this theory.

The idea of a universal theory does not exclusively belong to physicists
though. We mentioned already in Section 4 that Landau's paper [67] made a
big impact on the theory of singularities and catastrophes. Rene Thom-the
founding father of the theory of catastrophes-also was looking for the
theory of everything. \ In his book [122] he made an attempt to apply the
theory of singularities to biology. Later on Robert Gilmore using Thom's
results developed many applications of catastrophe theory to other
disciplines [123]. It is fair to say that the mathematical theory of
catastrophes is related to physical theory of random
fragmentation-coagulation processes as thermodynamics is related to
statistical mechanics. We believe, that it is this stochastic theory which
really deserves to be called the universal theory of everything. Hopefully,
without causing a global catastrophe, the millennium into which we are
entering is going to be the millennium of \ incredibly versatile succession
of hallmarks in the theory of universal catastrophes.

\ 

\textbf{Appendix A}. \textbf{Connections between the gamma and Dirichlet
distributions}

\bigskip\ 

Using results of our paper [83] especially, equation (3.27) of this
reference, such a connection can be easily established. Indeed, consider $%
n+1 $ independently \ distributed \ random gamma variables with exponents $%
\alpha _{1},...,\alpha _{n+1}$. The joint probability density for such
variables is given by 
\begin{equation}
p_{Y_{1}},..._{Y_{n+1}}(s_{1},...,s_{n+1})=\frac{1}{\Gamma (\alpha _{1})}%
\cdot \cdot \cdot \frac{1}{\Gamma (\alpha _{n+1})}s_{1}^{\alpha _{1}-1}\cdot
\cdot \cdot s_{n+1}^{\alpha _{n+1}-1}.  \tag{A.1}
\end{equation}%
Let now $s_{i}=t_{i}t,$ \ where $t_{i}$ are chosen in such a way that $%
\sum\nolimits_{n=1}^{n+1}t_{i}=1.$ Then, using such a substitution in (A.1)
we obtain at once: 
\begin{eqnarray}
p_{u_{1}},..._{u_{n+1}}(t_{1},...,t_{n+1}) &=&[\int\limits_{0}^{\infty
}t^{\alpha -1}e^{-t}]\frac{1}{\Gamma (\alpha _{1})}\cdot \cdot \cdot \frac{1%
}{\Gamma (\alpha _{n+1})}t_{1}^{\alpha _{1}-1}\cdot \cdot \cdot
t_{n+1}^{\alpha _{n+1}-1}\text{ , provided that }  \notag \\
1 &=&\sum\nolimits_{n=1}^{n+1}t_{i}.  \TCItag{A..2}
\end{eqnarray}%
Since $\alpha =\alpha _{1}+\cdot \cdot \cdot +\alpha _{n+1}$, we obtain: $%
\int\limits_{0}^{\infty }t^{\alpha -1}e^{-t}=\Gamma (\alpha _{1}+\cdot \cdot
\cdot +\alpha _{n+1}),$ so that the density of probability (A.2) is indeed
of Dirichlet-type\ given by (5.2).

\ 

\textbf{Appendix B. Some facts from combinatorics of the symmetric group }$%
S_{n}$

\ 

\ 

Suppose we have a finite set X. $\forall x\in \mathbf{X}$ consider a
bijection $\mathbf{X}\longrightarrow $X made of some permutation sequence: $%
x,$ $\pi (x),\pi ^{2}(x),...$Because the set is finite, we must have $\pi
^{m}(x)=x$ for some $m\geq 1.$ A sequence $(x,$ $\pi (x),\pi ^{2}(x),..,\pi
^{m-1}(x))=C_{m}$ is called a \textsl{cycle of length m}. The set $\mathbf{X}
$ can be subdivided into a \textsl{disjoint} product of cycles so that any
permutation $\pi $ is just a product of these cycles. Normally such a
product is not uniquely defined. To make it uniquely defined, we have to
assume that the set \textbf{X} is ordered according to a certain rule. The,
standard cycle representation can be constructed by requiring that: a) each
cycle is written with its largest element first, and b) the cycles are
written in increasing order of their respective largest elements. Let $N$ be
some integer and consider a decomposition of $N$ as $N=\sum%
\nolimits_{i=1}^{K}n_{i}$ . We say that $\ \mathit{n\equiv }%
(n_{0},...,n_{K}) $ is the \textsl{partition} of $N$ (or $n\vdash N).$The
same result can be achieved if, instead we would consider the following
decomposition of $N$: $N=\sum\nolimits_{i=1}^{N}ic_{i\text{ }}$ where,
according to our conventions, we have $c_{i}\equiv c_{i}(\pi )$ as the
number of cycles of length $i$. The total number of cycles then is given by $%
K$=$\sum\nolimits_{i=1}^{N}c_{i\text{ }}.$ Define a number $\tilde{S}(N,K)$
as the number of permutations of \textbf{X} with exactly $K$ cycles. Then,
the \textsl{Stirling} number of the \textsl{first }kind can be defined as $%
S(N,K):=(-1)^{N-K}\tilde{S}(N,K).$ The numbers $\tilde{S}(N,K)$ can be
obtained recursively using the following recurrence relation:%
\begin{equation}
\tilde{S}(N,K)=(N-1)\tilde{S}(N-1,K)+\tilde{S}(N-1,K-1),\text{ }N,K\geq 1;%
\text{ }\tilde{S}(0,0)=1.  \tag{B.1}
\end{equation}%
Use of this recurrence allows us to obtain the following important result:%
\begin{equation}
\sum\limits_{K=0}^{N}\tilde{S}(N,K)x^{K}=x(x+1)(x+2)\cdot \cdot \cdot
(x+N-1).  \tag{B.2}
\end{equation}%
Let now $x=1$ in $(B.2),$ then we can define the probability $p(K;N)=\tilde{S%
}(N,K)/N!$ Furthermore, one can define yet another probability by
introducing a notation $[x]^{N}=x(x+1)(x+2)\cdot \cdot \cdot (x+N-1).$ Then,
we obtain: 
\begin{equation}
\sum\limits_{K=0}^{N}\tilde{S}(N,K)\frac{x^{K}}{[x]^{N}}=\sum%
\limits_{K=0}^{N}P_{K}(N;x)=1.  \tag{B.3}
\end{equation}%
Such defined probability $P_{K}(N;x)$ can be further rewritten in view of
the famous result by Cauchy. To obtain his result, we introduce the
generating function%
\begin{equation}
\mathcal{F}_{K}^{N}(\mathbf{x)=}\sum\limits_{_{\substack{ %
K=\sum\nolimits_{i=1}^{N}c_{i\text{ }}  \\ N=\sum\nolimits_{i=1}^{N}ic_{i%
\text{ }}}}}\frac{N!}{1^{c_{1}}c_{1}!2^{c_{2}}c_{2}!\cdot \cdot \cdot
N^{c_{N}}c_{N}!}x_{1}^{c_{1}}\cdot \cdot \cdot x_{N}^{c_{N}}  \tag{B.4a}
\end{equation}%
and require that $\tilde{S}(N,K)\mathbf{x}^{K}=\mathcal{F}_{K}^{N}(\mathbf{x)%
}$. This can happen only if 
\begin{equation}
\sum\limits_{K=0}^{N}\sum\limits_{_{\substack{ K=\sum\nolimits_{i=1}^{N}c_{i%
\text{ }}  \\ N=\sum\nolimits_{i=1}^{N}ic_{i\text{ }}}}}\frac{N!}{%
1^{c_{1}}c_{1}!2^{c_{2}}c_{2}!\cdot \cdot \cdot N^{c_{N}}c_{N}!}=1 
\tag{B.4b}
\end{equation}%
Thus, we obtain: 
\begin{equation}
\tilde{S}(N,K)=\prod\limits_{i=1}^{K}\frac{N!}{i^{c_{i}}c_{i}!}\text{ , \
provided that }K=\sum\nolimits_{i=1}^{N}c_{i\text{ }}\text{ and }%
N=\sum\nolimits_{i=1}^{N}ic.  \tag{B.5}
\end{equation}%
In these notations the Ewens sampling formula acquires the following
canonical form$:$%
\begin{equation}
P_{K}(N;x)\equiv \frac{x^{K}}{[x]^{N}}\prod\limits_{i=1}^{K}\frac{N!}{%
i^{c_{i}}c_{i}!}\text{ ,}\ \text{provided that }K=\sum\nolimits_{i=1}^{N}c_{i%
\text{ }}\text{and }N=\sum\nolimits_{i=1}^{N}ic.  \tag{B.6}
\end{equation}

\bigskip

\bigskip

{\large References}

\ 

[1] \ L.Landau, "On fundamental problems", Collected Works.

\ \ \ \ \ Vol.2, p. 421 (Nauka, Moscow, 1969)\footnote{%
This reference is taken from the original Russian Landau's edition of his
collected works. There is an English translation \ edited by D.Ter Haar:
"Collected works by L.D.Landau (Gordon and Breach, New York, 1967). To avoid
any ambiguities, we decided to use the original Russian edition as reference
here and below.}

[2] \ E.Fermi, "Angular distribution of the pions in high energy

\ \ \ \ \ nuclear collisions", Phys.Rev. \textbf{81} 683 (1951).

[3] \ W.Heisenberg, "Quantum theory of fields and elementary

\ \ \ \ \ particles", Rev.Mod Phys. \textbf{29}, 269 (1957).\ 

[4] \ M.Tannenbaum, " Recent results in relativistic heavy ion

\ \ \ \ \ collisions: from 'a new state of matter' to 'the perfect fluid'",

\ \ \ \ \ Reports on Progrress on Phys. \textbf{69}, 2005 (2006).

[5] \ D.Gross and F.Wilczek, "Ultraviolet behavior of non-Abelian

\ \ \ \ \ gauge theories", Phys.Rev.Lett. \textbf{30} 1343 (1973).

[6] \ H.D Politzer, "Reliable perturbative results for strong interactions",

\ \ \ \ \ Phys.Rev.Lett. \textbf{30} , 1346 (1973).

[7] \ C.Yang and R.Mills, "Conservation of isotopic spin and isotopic

\ \ \ \ \ gauge invariance", Phys.Rev.\textbf{96}, 19 (1954).

[8] \ L.Faddeev and V.Popov, "Feynman diagrams for the Yang-Mills

\ \ \ \ \ field", Phys.Lett.B \textbf{25}, 29 (1967).

[9] \ P.Dirac, "Generalized Hamiltonian dynamics",

\ \ \ \ \ Canadian J.Math. \textbf{2,} 129 (1950).

[10] L.Faddeev and A.Slavnov, Gauge Fields. Introduction to

\ \ \ \ \ \ Quantum Theory (Benjamin/Cummings, Reading, MA,1980).

[11] M.Veltman, Diagramatica: A Path to Feynman Rules

\ \ \ \ \ \ (Cambridge University Press, Cambridge, UK, 1994).

[12] J.Bjorken, "Asymptotic sum rules at infinite momentum",

\ \ \ \ \ \ Phys.Rev\textbf{179}, 1547 (1969).

[13] G.Veneziano,"Construction of a crossing-symmetric,

\ \ \ \ \ \ Regge-behaved amplitude for linearly rising trajectories",

\ \ \ \ \ \ Nuovo Cimento\textbf{17A}, 190 (1968).

[14] P.Frampton, Dual Resonance Models

\ \ \ \ \ \ ( Benjamin/Cummings, Reading, MA, 1974).

[15] E. Kiritsis, String Theory in a Nutshell

\ \ \ \ \ \ (Princeton University Press, Princeton, 2007).

[16] C. Schubert, "Perturbative quantum field theory in the

\ \ \ \ \ \ string-inspired formalism", Phys.Reports \textbf{355},73 (2001).

[17] J.Goldstone, "Field theories with superconductor solutions",

\ \ \ \ \ \ Nuovo Cimento \textbf{19}, 154 (1961).

[18] P.Higgs, "Spontaneous symmetry breakdown without massless bosons,"

\ \ \ \ \ \ Phys.Rev. \textbf{145}, 1156 (1966).

[19] \ L.Landau, "About theory of phase transitions", Collected Works.

\ \ \ \ \ \ \ Vol.1, p.234 (Nauka, Moscow, 1969).

[20] \ L.Landau, "About theory of superconductivity",

\ \ \ \ \ \ together with V.Ginzburg, Collected Works.

\ \ \ \ \ \ Vol.2, p.126 (Nauka, Moscow, 1969).

[21] \ D.Bardin and G. Passarino, The Standard Model in the Making

\ \ \ \ \ \ \ (Clarendon Press, Oxford, 1999).

[22] \ \ W.Cottingham and D. Greenwood, An Introduction to the

\ \ \ \ \ \ \ Standard Model of Particle Physics

\ \ \ \ \ \ \ (Cambridge University Press, Cambridge 2007).

[23] \ \ R.Utiyama, " Invariant- theoretical interpretation of interaction",

\ \ \ \ \ \ \ Phys.Rev.\textbf{101}, 1597 (1956).

[24] \ A.Kholodenko and E.Ballard, "From Ginzburg-Landau to

\ \ \ \ \ \ \ Hilbert-Einstein via Yamabe", Physica \textbf{A380},115 (2007).

[25] \ A.Kholodenko, "Towards physically motivated proofs of the

\ \ \ \ \ \ \ Poincare$^{\prime }$ and geometrization conjectures",

\ \ \ \ \ \ \ J.Geom.Phys.\textbf{58}, 259 (2008).

[26] \ \ G.Lisi," An exceptionally simple theory of everything", arxiv:
0711.0770.

[27] \ \ A.Zamomolodchikov and Al.Zamolodchikov, "Factorized S-Matrices

\ \ \ \ \ \ \ \ in two dimensions as the exact solutions of certain
relativistic

\ \ \ \ \ \ \ \ quantum field theory models", Ann.Phys. 1\textbf{29}, 253
(1979).

[28] \ \ J.McGuire,"Study of exactly soluble one-dimensional

\ \ \ \ \ \ \ N-body problems", J.Math.Phys.\textbf{5}, 622 (1964).

\bigskip \lbrack 29] \ C.N.Yang, "Some exact results for the many-body
problem in one

\ \ \ \ \ \ \ dimension with repulsive delta-function interaction",

\ \ \ \ \ \ \ Phys.Rev.Lett.\textbf{19},1312 (1967).

[30] \ R.Baxter, "Partition function of the eight-vertex lattice model",

\ \ \ \ \ \ \ Ann.Phys.\textbf{70},193 (1972).

[31] \ M.Jimbo,Yang-Baxter Equation in Integrable Systems

\ \ \ \ \ \ \ (World Scientific, Singapore, 1990).

[32] \ C.Gomez, M.Ruiz-Altaba and G. Sierra, Quantum Groups

\ \ \ \ \ \ in Two-Dimensional Physics

\ \ \ \ \ \ (Cambridge University Press, Cambridge 1996).

[33] \ A.Belavin and V.Drinfeld, "Solutions of the cclassical Yang-Baxter

\ \ \ \ \ \ \ equation for simple Lie algebras", funct.Anal.Appl. \textbf{16}%
,150 (1983).

[34] \ V.Drinfeld, "Hopf algebras and the quantum Yang-Baxter equation",

\ \ \ \ \ \ \ Sov.Math.Doclady \textbf{32}, 254 (1985).

[35] \ C.Kassel, Quantum Groups (Springer-Verlag, Berlin, 1995).

[36] \ V.Chari and A.Pressley, Quantum Groups

\ \ \ \ \ \ \ (Cambridge University Press, Cambridge 1995).

[37] \ A.Kholodenko, "Heinsenberg honeycombs solve Veneziano puzzle"

\ \ \ \ \ \ \ \ arxiv.hep-th/0608117.

[38] \ V.Knizhnik and A.Zamolodchikov, " Current algebra and Wess-Zumino

\ \ \ \ \ \ \ model in two dimensions", Nucl.Phys.\textbf{B247}, 83 (1984).

[39] \ A.Atland and B.Simons, Condensed Matter Field Theory

\ \ \ \ \ \ \ (Cambridge University Press, Cambridge 2006).

[40] \ E.Witten, 'Quantum field theory and the Jones polynomial",

\ \ \ \ \ \ Comm.Math.Phys.\textbf{121},351 (1989).

[41] \ E.Witten, "On holomorphic factorization of WZW and coset models",

\ \ \ \ \ \ \ Comm.Math.Phys. \textbf{144},189 (1992).

[42] \ A.Tsvelik, Quantum Field Theory in Condensed Matter Physics

\ \ \ \ \ \ \ \ (Cambridge University Press, Cambridge 2003).

[43] \ P.Di Francesco, P.Mathieu and D.Senechal, Conformal Field Theory,

\ \ \ \ \ \ \ (Springer-Verlag, Berlin, 1997).

[44] \ A.Varchenko, Special Functions, KZ-Type Equations and

\ \ \ \ \ \ \ Representation Theory (AMS Publishers, Providence, RI, 2003).

[45] \ \ V.Golubeva, "Some problems in analytic theory of Feynman's

\ \ \ \ \ \ \ integrals", Russian Math Surveys \textbf{31}, 139 (1976).

[46] \ A.Kholodenko, "Quantum signatures of Solar System dynamics",

\ \ \ \ \ \ \ arxiv:0707.3992.

[47] \ A.Kholodenko, " New strings for old Veneziano amplitudes IV.

\ \ \ \ \ \ \ Connections with spin chains and other stochastic systems",

\ \ \ \ \ \ \ arxiv: 0805.0113.

[48] \ \ M.Atiyah and R.Bott, " The Yang-Mills equation over Riemann

\ \ \ \ \ \ \ \ surfaces", Phil.Trans. R.Soc.London \textbf{A 308}, 523
(1982).

[49] \ \ E.Witten, "On quantum gauge theories in two dimensions",

\ \ \ \ \ \ \ \ Comm.Math.Phys. \textbf{141}, 153 (1991).

[50] \ \ E.Witten, "Two dimensional gauge theories revisited",

\ \ \ \ \ \ \ \ J.Geom..Phys.\textbf{9}, 303 (1992).

[51] \ \ A. Migdal, "Recursion equations in gauge theories",

\ \ \ \ \ \ \ \ Sov.Phys.JETP \textbf{42}, 413 (1975).

[52] \ \ A.Belavin, A.Polyakov, A.Schwartz and Yu.Tupkin,

\ \ \ \ \ \ \ "Pseudoparticle solution of the Yang-Mills equations",

\ \ \ \ \ \ \ \ Phys.Lett. \textbf{B 59}, 85 (1975).

[53] \ \ G. 't Hooft, "Computations of the quantum effects due

\ \ \ \ \ \ \ \ to a four-dimensional pseudoparticle", Phys.Rev.\textbf{D14}%
,3432 (1976).

[54] \ \ A.Belavin, "The inverse scattering problem and instanton

\ \ \ \ \ \ \ \ construction by algebraic geometry", Soviet Sci.Reviews,
Section A:

\ \ \ \ \ \ \ \ Physics.Reviews \textbf{1}, 1 (1979).

[55] \ \ S.Donaldson and P.Kronheimer, The Geometry of Four-Manifolds

\ \ \ \ \ \ \ \ (Clarendon Press, Oxford, 1990).

[56] \ \ Yu. Soloviyov, "Topology of four-manifolds",

\ \ \ \ \ \ \ \ Russian Math.Surveys \textbf{46}, 145 (1991).

[57] \ \ \ N.Tyurin, "Instantons and monopoles" ,

\ \ \ \ \ \ \ \ Russian Math.Surveys \textbf{57}, 305 (2002).

[58] \ \ L.Nicolaescu, Notes on Seiberg-Witten Theory

\ \ \ \ \ \ \ \ (AMS Publishers, Providence, RI, 2000).

[59] \ \ J.Jost, Riemannian Geometry and Geometric Analysis

\ \ \ \ \ \ \ \ (Springer-Verlag, Berlin, 2005).

[60] \ \ L.Pismen, Vortices in Nonlinear Fields

\ \ \ \ \ \ \ \ (Clarendon Press, Oxford, 1999)

[61] \ \ Y.Yang, Solitons in Field Theory and Nonlinear Analysis

\ \ \ \ \ \ \ \ (Springer-Verlag, Berlin, 2001).

[62] \ \ L.Mason and N. Woodhouse, Integrability,

\ \ \ \ \ \ \ \ Self-Duality, and Twistor Theory

\ \ \ \ \ \ \ \ (Clarendon Press, Oxford, 1996).

[63] \ \ A.Marshakov, Seiberg-Witten Theory and Integrable Systems

\ \ \ \ \ \ \ \ (World Scientific, Singapore, 1999).

[64] \ \ M.Dunajsko, L.Mason and N.Woodhouse, " From 2d integrable

\ \ \ \ \ \ \ \ systems to self-dual gravity", J.Phys.A 31, 6019 (1998).

[65] \ \ E.Witten, "Mirror symmetry, Hitchin's equations and

\ \ \ \ \ \ \ \ Langlands duality", arxiv: 0802.0999.

[66] \ \ J. J.Zinn-Justin, Quantum Field Theory and Critical Phenomena

\ \ \ \ \ \ \ \ (Clarendon Press, Oxford, 1997).

[67] \ \ L.Landau,"On analytical properties of the vertex parts in

\ \ \ \ \ \ \ \ quantum field theory", Collected Works.

\ \ \ \ \ \ \ \ Vol.2, p. 406 (Nauka, Moscow, 1969).

[68] \ \ P.Orlik and H.Terrao, Arrangements and Hypergeometric Integrals

\ \ \ \ \ \ \ \ (MSJ Memoirs, Tokyo, 2001).

[69] \ \ M.Yoshida, Hypergeometric Functions, My Love

\ \ \ \ \ \ \ \ (Vieweg\&Sohn Co., Wiesbaden, 1997).

[70] \ \ \ M.Saito, B.Sturmfels and N.Takayama, Gr\"{o}bner Deformations

\ \ \ \ \ \ \ \ \ of Hypergeometric Differential Equations

\ \ \ \ \ \ \ \ \ (Springer-Verlag, Berlin, 2000).

[71] \ \ \ J.Bjorken and S.Drell, Relativistic quantum Fields, Vol.2

\ \ \ \ \ \ \ \ \ (McGraw Hill Co., NY, 1965).

[72] \ \ \ N.Reshetikhin and P.Wiegmann," Towards classification of

\ \ \ \ \ \ \ \ \ completely integrable quantum field theories," Phys.Lett.%
\textbf{B189},

\ \ \ \ \ \ \ \ \ 125 (1987).

[73] \ \ \ A.Connes, D. Kreimer,\ "Renormalization in quantum

\ \ \ \ \ \ \ \ \ field theory and the Riemann-Hilbert problem I: the Hopf

\ \ \ \ \ \ \ \ \ algebra structure of graphs and the main theorem,"

\ \ \ \ \ \ \ \ \ Commun.Math.Phys. \textbf{210}, 249 (2000).

[74] \ \ \ A.Connes, D. Kreimer,\ "Renormalization in quantum

\ \ \ \ \ \ \ \ \ field theory and the Riemann-Hilbert problem I: the beta
function,

\ \ \ \ \ \ \ \ \ diffeomorphism and the renormalization group,"

\ \ \ \ \ \ \ \ \ Commun.Math.Phys. \textbf{216}, 215 (2001).

[75] \ \ \ \ A.Connes, M.Marcolli, 'Quantum fields and motives,"

\ \ \ \ \ \ \ \ \ J.Geom.Phys.\textbf{56}, 55 (2006).

[76] \ \ \ \ S.Bloch, D.Kreimer, "Mixed Hodge structures and renormalization

\ \ \ \ \ \ \ \ \ \ in physics," arxiv: 0804.4399.

[77] \ \ \ M.Atiyah, "The impact of Thom's cobordism theory,"

\ \ \ \ \ \ \ \ \ BAMS \textbf{41}, 337 (2004).

[78] \ \ \ F.Pham, Introduction A L'Etude Topologique des

\ \ \ \ \ \ \ \ \ \ Singularities De Landau (Gauthier-Villars, Paris, 1967).

[79] \ \ \ \ F.Pham, "Formules de Picard-Lefschetz generalisees

\ \ \ \ \ \ \ \ \ \ et ramification des integrales,"

\ \ \ \ \ \ \ \ \ \ Bull.Soc.Math.France \textbf{93},333 (1965).

[80] \ \ \ \ D.Fotiadi, M.Froissart, J.Lascoux and F.Pham,

\ \ \ \ \ \ \ \ \ " Application of the theorem about isotopy,"

\ \ \ \ \ \ \ \ \ \ Topology \textbf{4}, 159 (1965).

[81] \ \ \ \ V.Arnol'd, S.Gussein-Zade and A.Varchenko,

\ \ \ \ \ \ \ \ \ \ Singularities of Differentiable Maps, Vol.2

\ \ \ \ \ \ \ \ \ \ (Birkhauser, Boston, 1988).

[82]\ \ \ \ \ J.Carlson, S.Muller-Stach and C.Peters,

\ \ \ \ \ \ \ \ \ \ Period Mappings and Period Domains

\ \ \ \ \ \ \ \ \ (Cambridge University Press, Cambridge, 2003).

[83] \ \ \ A.Kholodenko, "New strings for old Veneziano

\ \ \ \ \ \ \ \ \ amplitudes I. Analytical tratment,"

\ \ \ \ \ \ \ \ \ J.Geom.Phys. \textbf{55}, 50 (2005).

[84] \ \ \ A.Kholodenko, "New strings for old Veneziano

\ \ \ \ \ \ \ \ \ amplitudes II. Group-theoretic treatment,"

\ \ \ \ \ \ \ \ \ J.Geom.Physics \textbf{56}, 1387 (2006).

[85] \ \ \ \ A.Kholodenko, "New strings for old Veneziano amplitudes III.

\ \ \ \ \ \ \ \ \ \ Symplectic treatment," J.Geom.Phys. \textbf{56}, 1433
(2006).

[86] \ \ \ \ R.Eden, P.Landshoff, D.Olive and J.Polkinghorne,

\ \ \ \ \ \ \ \ \ \ The Analytic S-Matrix,

\ \ \ \ \ \ \ \ \ \ (Cambridge University Press, Cambridge, 1966).

[87] \ \ \ \ P.Collins, An Introduction to Regge Theory and

\ \ \ \ \ \ \ \ \ \ High Energy Physics (Cambridge University Press,

\ \ \ \ \ \ \ \ \ \ Cambridge, 1977).

[88] \ \ \ \ J.Collins, Renormalization

\ \ \ \ \ \ \ \ \ \ (Cambridge University Press, Cambridge, 1984).

[89] \ \ \ \ C.Bogner, S.Weinzezierl, " Periods and feynman integrals,"

\ \ \ \ \ \ \ \ \ \ arxiv:0711.4863

[90] \ \ \ \ \ S. Wienzierl, Feynman integrals and multiple polylogaritms,"

\ \ \ \ \ \ \ \ \ \ \ arxiv: 0707.0900

[91] \ \ \ \ \ O.Tarasov, "Hypergeometric representation of the two-loop

\ \ \ \ \ \ \ \ \ \ \ equal mass sunrize diagram," arxiv:hep-ph/0603227.

[92] \ \ \ \ \ A.Smirnov and V.Smirnov," On \ the reduction of Feynman

\ \ \ \ \ \ \ \ \ \ \ integrals to master integrals," arxiv: 0707.3993.

[93] \ \ \ \ \ B.Shabat, Introduction to Complex Analysis, Part II :

\ \ \ \ \ \ \ \ \ \ Functions of Several Complex Variables

\ \ \ \ \ \ \ \ \ \ (AMS, Providence, RI, 1992).

[94] \ \ \ \ P.Griffiths, "On periods of certain rational integrals,"

\ \ \ \ \ \ \ \ \ \ Ann.Math. \textbf{90}, 460 (1969).

[95] \ \ \ \ A.Kholodenko, "Traces of mirror symmetry in Nature,"

\ \ \ \ \ \ \ \ \ International Mathematical.Forum \textbf{3},151 (2008).

[96] \ \ \ I.Gelfand, M.Kapranov and A.Zelevinsky, "Generalized

\ \ \ \ \ \ \ \ \ Euler integrals and A-hypergeometric functions,"

\ \ \ \ \ \ \ \ \ Advances in Mathematics \textbf{84}, 255 (1990).

[97] \ \ \ J. Sienstra," GKZ hypergeometric structures",

\ \ \ \ \ \ \ \ \ arxiv: math. AG/0511351.

[98] \ \ \ D.Kox and S.Katz, Mirror Symmetry and

\ \ \ \ \ \ \ \ \ Algebraic Geometry

\ \ \ \ \ \ \ \ \ (AMS Publishers, Providence, RI 1999).

[99] \ \ \ \ I.Gelfand, A.Zelevinskii and M.Kapranov,

\ \ \ \ \ \ \ \ \ "Hypergeometric functions and toric manifolds",

\ \ \ \ \ \ \ \ \ \ Functional Analysis and Appl. \textbf{23}, 12 (1989).

[100] \ \ \ A.Kholodenko, "Development of new apartment

\ \ \ \ \ \ \ \ \ \ buildings for strings and conformal field theories"

\ \ \ \ \ \ \ \ \ \ in New Developments in String Theory Research.

\ \ \ \ \ \ \ \ \ \ pages 1-83 (Nova \ Sci. Publ. Inc, New York, 2006).

[101] \ \ \ H.Bateman and A.Erdelyi, Higher Transcendental

\ \ \ \ \ \ \ \ \ \ Functions, Vol.1 (McGraw Hill, New York, 1953).

[102] \ \ \ B.Schroer,"String theory and the crisis of particle physics,"

\ \ \ \ \ \ \ \ \ \ arxiv:08051911.

[103] \ \ \ J.Bertoin, Random Fragmentation and Coagulation

\ \ \ \ \ \ \ \ \ \ \ Processes (Cambridge University Press, Cambridge,
2006).

[104] \ \ \ J.Pitman, Combinatorial Stochastic Processes

\ \ \ \ \ \ \ \ \ \ \ (Springer-Verlag, Berlin, 2006).

[105] \ \ \ \ R.Arratia,A.Barbour and S.Tavare, Logarithmic

\ \ \ \ \ \ \ \ \ \ \ Combinatorial Structures: a Probabilistic Approach

\ \ \ \ \ \ \ \ \ \ \ (European Mathematical Society, Zurich, 2003).

[106] \ \ \ \ A. Mekjian, "Model for studying branching processes,

\ \ \ \ \ \ \ \ \ \ \ \ multiplicity distribution and non-Poissonian
fluctuations

\ \ \ \ \ \ \ \ \ \ \ \ in heavy-ion collisions", PRL \textbf{86}, 220
(2001).

[107] \ \ \ \ \ N.Balakrishnan and V.Nevzorov,

\ \ \ \ \ \ \ \ \ \ \ \ A Primer on Statistical Distributions

\ \ \ \ \ \ \ \ \ \ \ \ (John Wiley \&Sons, Inc. Hoboken, NJ, 2003).

[108] \ \ \ \ \ J.Kingman, Poisson Processes

\ \ \ \ \ \ \ \ \ \ \ \ (Clarendon Press, Oxford, 1993).

[109] \ \ \ \ \ A.Mekjian, "Cluster \ distribution in physics

\ \ \ \ \ \ \ \ \ \ \ \ and genetic diversity", Phys.Rev.\textbf{A 44}, 8361
(1991).

[110] \ \ \ \ \ W.Ewens, Mathematical Populational Genetics,

\ \ \ \ \ \ \ \ \ \ \ \ (Springer-Verlag, Berlin, 2004).

[111] \ \ \ \ \ G.Watterson, "The stationary distribution of the

\ \ \ \ \ \ \ \ \ \ \ \ infinitely-many neutral alleles:diffusion model",

\ \ \ \ \ \ \ \ \ \ \ \ J.Appl, Probability \textbf{13}, 639 (1976).

[112] \ \ \ \ \ M.Aoki, "Open models of share markets with

\ \ \ \ \ \ \ \ \ \ \ \ two dominant types of participants",

\ \ \ \ \ \ \ \ \ \ \ \ J.of Economic Behaviour\&Organization \textbf{49},
199 (2002).

[113] \ \ \ \ \ J.Kingman,"The population structure associated

\ \ \ \ \ \ \ \ \ \ \ \ with the Evens sampling formula", Theoret.Pop.

\ \ \ \ \ \ \ \ \ \ \ \ Biology \textbf{11}, 274 (1977).

[114] \ \ \ \ \ H.Harari,"Duality diagrams", PRL \textbf{22}, 562 (1969).

[115] \ \ \ \ \ D.Leach, Genetic Recombination

\ \ \ \ \ \ \ \ \ \ \ \ (Blackwell Science Ltd.Oxford, UK, 1996).

[116] \ \ \ \ \ P.Freund,"Two component duality and strings",

\ \ \ \ \ \ \ \ \ \ \ \ arxiv: 0708.1983

[117] \ \ \ \ \ S.Mirkin, "Structure and biology of H DNA" in

\ \ \ \ \ \ \ \ \ \ \ \ Triple Helix Forming \ Oligonucleotides, p195-222

\ \ \ \ \ \ \ \ \ \ \ \ (Kluwer Academic, Boston, 1999).

[118] \ \ \ \ \ \ D.McQuarrie, "Stochastic approach to chemical kinetics,

\ \ \ \ \ \ \ \ \ \ \ \ \ J.Appl.Prob. \textbf{4}, 413 (1967).

[119] \ \ \ \ \ \ I.Darvey, B.Ninham and P.Staff, "Stochastic models

\ \ \ \ \ \ \ \ \ \ \ \ \ for second -order chemical reaction kinetics.

\ \ \ \ \ \ \ \ \ \ \ \ \ The Equilibrium state", J.Chem.Phys. \textbf{45},
2145 (1966).

[120] \ \ \ \ \ \ J.Kingman,"Dynamics of neutral mutation",

\ \ \ \ \ \ \ \ \ \ \ \ \ Proc.Roy.Soc.London \textbf{A 363}, 135 (1978).

[121] \ \ \ \ \ \ G.Felder and A.Veselov, Polynomial solutions

\ \ \ \ \ \ \ \ \ \ \ \ \ of the Knizhnik-Zamolodchikov equations and

\ \ \ \ \ \ \ \ \ \ \ \ \ Schur-Weyl duality", Intern.math.Res. Notices

\ \ \ \ \ \ \ \ \ \ \ \ \ Article ID rnm046 (2007).

[122] \ \ \ \ \ \ R.Thom, Structural Stability and Morphogenesis

\ \ \ \ \ \ \ \ \ \ \ \ \ (W.A.Benjamin, Inc. Reading, MA, 1975).

[123] \ \ \ \ \ \ R.Gilmore, Catastrophe Theory for Scientists

\ \ \ \ \ \ \ \ \ \ \ \ \ and Engineers (John Wiley \& Sons, NY, 1981).

\ \ \ 

\ \ \ \ \ 

\ \ \ 

\bigskip

\bigskip

\bigskip

\bigskip

\bigskip

\bigskip

\bigskip

\end{document}